\RequirePackage[displaymath]{lineno} 
\documentclass[aps,prd,twocolumn,showpacs,amsmath,amssymb]{revtex4-1}
\usepackage{epsfig}
\usepackage{graphicx}
\usepackage{dcolumn}
\usepackage{bm}
\usepackage{ltablex,booktabs}
\usepackage{overpic}
\usepackage{subfigure}
\usepackage{float}
\usepackage{color}
\usepackage{amsmath}
\usepackage{mathcomp}
\usepackage{mathrsfs}
\usepackage{multirow}
\usepackage{rotating}
\usepackage{amssymb}
\usepackage{gensymb}
\usepackage{amsmath}
\usepackage{tabularx}

\newcolumntype{L}{D{$\pm$}{$\pm$}{6,4}}
\begin{document}
\normalsize
\parskip=5pt plus 1pt minus 1pt

\title{ \boldmath Amplitude analysis and branching fraction measurement of $D_{s}^{+} \rightarrow K^{+}K^{-}\pi^{+}$ }

\author{
		\begin{center}
	M.~Ablikim$^{1}$, M.~N.~Achasov$^{10,c}$, P.~Adlarson$^{64}$, S. ~Ahmed$^{15}$, M.~Albrecht$^{4}$, A.~Amoroso$^{63A,63C}$, Q.~An$^{60,48}$, ~Anita$^{21}$, Y.~Bai$^{47}$, O.~Bakina$^{29}$
% ...
	\\
		\vspace{0.2cm}
		(BESIII Collaboration)\\
		\vspace{0.2cm} {\it
	$^{1}$ Institute of High Energy Physics, Beijing 100049, People's Republic of China\\
% ...
$^{4}$ Bochum  Ruhr-University, D-44780 Bochum, Germany\\
% ...
$^{10}$ G.I. Budker Institute of Nuclear Physics SB RAS (BINP), Novosibirsk 630090, Russia\\
% ...
$^{15}$ Helmholtz Institute Mainz, Johann-Joachim-Becher-Weg 45, D-55099 Mainz, Germany\\
% ...
$^{21}$ Indian Institute of Technology Madras, Chennai 600036, India\\
% ...
$^{29}$ Joint Institute for Nuclear Research, 141980 Dubna, Moscow region, Russia\\
% ...
$^{47}$ Southeast University, Nanjing 211100, People's Republic of China\\$^{48}$ State Key Laboratory of Particle Detection and Electronics, Beijing 100049, Hefei 230026, People's Republic of China\\
% ...
$^{60}$ University of Science and Technology of China, Hefei 230026, People's Republic of China\\
% ...
$^{63}$ (A)University of Turin, I-10125, Turin, Italy; (B)University of Eastern Piedmont, I-15121, Alessandria, Italy; (C)INFN, I-10125, Turin, Italy\\$^{64}$ Uppsala University, Box 516, SE-75120 Uppsala, Sweden\\
% ...
$^{c}$ Also at the Novosibirsk State University, Novosibirsk, 630090, Russia\\
% ...
		}\end{center}
		\vspace{0.4cm}
}

\affiliation{}
\date{\today}
\begin{abstract}
  We report an amplitude analysis and branching fraction measurement of $D_{s}^{+} \rightarrow K^{+}K^{-}\pi^{+}$ decay using a data sample of 3.19 $\rm fb^{-1}$ recorded with BESIII detector at a center-of-mass energy of 4.178 GeV. 
  We perform a model-independent partial wave analysis in the low $K^{+}K^{-}$ mass region to determine the $K^{+}K^{-}$ S-wave lineshape, 
  followed by an amplitude analysis of our very pure high-statistics sample.  
  With the detection efficiency based on the amplitude analysis results,
  the absolute branching fraction is measured to be
  ${\cal B}(D_{s}^{+} \rightarrow K^{+}K^{-}\pi^{+}) = (5.47\pm0.08_{{\rm stat}}\pm0.13_{{\rm sys}})\%$. 
\end{abstract}
\maketitle


\section{Introduction}
\label{sec:introduction}
The decay $D_{s}^{+} \rightarrow K^{+}K^{-}\pi^{+}$ is widely used as a 
reference mode in $D_{s}^{\pm}$ analyses because of its large branching 
fraction (BF) and low background contamination.
An amplitude analysis can reveal the intermediate states involved in this decay
and thereby reduce the detection efficiency systematic uncertainties.
The improved precision of the BF is important for $D_{s}^{\pm}$ analysis using this decay as a reference channel.
Furthermore, theoretical studies~\cite{PRD93-114010} predict the BFs of $D_{s}^{+} \rightarrow \bar{K}^{*}(892)^{0}K^{+}$ and $D_{s}^{+} \rightarrow \phi(1020)\pi^{+}$ to be in the range of $(3.9 - 4.2)\%$ and $(3.4 -  4.51)\%$, respectively.
Combining the results of the amplitude analysis and the BF measurement, 
one can obtain the BFs of such intermediate processes, 
which can help to improve the theoretical model~\cite{PRD93-114010}.
    
Dalitz plot analyses of the $D_{s}^{+} \rightarrow K^{+}K^{-}\pi^{+}$ decay have been performed by the E687~\cite{E687RES}, CLEO~\cite{2009CLEO} and BaBar~\cite{2011BABAR} collaborations.
    The E687 collaboration used about 700 pure signal events and did not take the 
$f_{0}(1370)\pi^{+}$ intermediate state into account. 
    In the CLEO analysis about 14400 events with a purity of 84.9\% 
were selected in an untagged analysis of 0.586 fb$^{-1}$ of data similar 
to the present analysis.  
    The analysis of BaBar collaboration used about 100\,000 events 
with a purity of about 95\%. 
    Table~\ref{PreviousAnalyses} shows the comparison of the fit 
fractions (FFs) from these previous Dalitz plot analyses.
\begin{table*}[htbp]
    \caption{Comparison of FFs for different decay modes.
        The first and second uncertainties are statistical and systematic, respectively.
    }
    \renewcommand\arraystretch{1.2}
    \label{PreviousAnalyses}
    \begin{center}
        \begin{tabular}{l|ccc}
            \hline\hline
            \multirow{2}{*}{Decay mode }&\multicolumn{3}{c}{FF(\%)}\cr 
            & \multicolumn{1}{c}{E687} &  \multicolumn{1}{c}{CLEO}  &\multicolumn{1}{c}{BaBar}   \\
            \hline
            $D_{s}^{+} \rightarrow \bar{K}^{*}(892)^{0}K^{+}$          & 47.8$\pm$4.6$\pm$4.0   & \hphantom{0}47.4$\pm$1.5$\pm$0.4  & \hphantom{0}47.9$\pm$0.5$\pm$0.5 \\
            $D_{s}^{+} \rightarrow \phi(1020)\pi^{+}$                  & 39.6$\pm$3.3$\pm$4.7   & \hphantom{0}42.2$\pm$1.6$\pm$0.3  & \hphantom{0}41.4$\pm$0.8$\pm$0.5 \\
            $D_{s}^{+} \rightarrow f_{0}(980)\pi^{+}$                  & 11.0$\pm$3.5$\pm$2.6   & \hphantom{0}28.2$\pm$1.9$\pm$1.8  & \hphantom{0}16.4$\pm$0.7$\pm$2.0 \\
            $D_{s}^{+} \rightarrow \bar{K}^{*}_{0}(1430)^{0}K^{+}$     & \hphantom{0}9.3$\pm$3.2$\pm$3.2   &  \hphantom{00}3.9$\pm$0.5$\pm$0.5  &  \hphantom{00}2.4$\pm$0.3$\pm$1.0  \\
            $D_{s}^{+} \rightarrow f_{0}(1710)\pi^{+}$                 & \hphantom{0}3.4$\pm$2.3$\pm$3.5   &  \hphantom{00}3.4$\pm$0.5$\pm$0.3  &  \hphantom{00}1.1$\pm$0.1$\pm$0.1  \\
            $D_{s}^{+} \rightarrow f_{0}(1370)\pi^{+}$                 &                        &  \hphantom{00}4.3$\pm$0.6$\pm$0.5&  \hphantom{00}1.1$\pm$0.1$\pm$0.2 \\ 
            $\begin{matrix}\sum {\rm FF}(\%)\end{matrix}$              &  111.1                    &129.5$\pm$4.4$\pm$2.0& 110.2$\pm$0.6$\pm$2.0  \\
                Events                                                 &  701$\pm$36               &  12226$\pm$122       &   96307$\pm$369        \\ 
                \hline\hline
            \end{tabular}
        \end{center}
    \end{table*}
    There are obvious differences between FFs of BaBar collaboration and CLEO collaboration. 
    
    The decay $D_{s}^{+} \rightarrow a_{0}(980)^{0}\pi^{+}$ has been observed through $D_{s}^{+} \rightarrow \pi^{+}\pi^{0}\eta$~\cite{pipi0eta},
    and should also be present in $D_{s}^{+} \rightarrow K^{+}K^{-}\pi^{+}$,
    which was not taken into account before.
    Due to the large overlap of $a_{0}(980) \rightarrow K^{+}K^{-}$ and $f_{0}(980) \rightarrow K^{+}K^{-}$ and their common $J^{PC}$, 
    we do not distinguish between them in this paper and denote the combined state as $S(980)$.
    A model-independent partial wave analysis (MIPWA) is performed to study this low-mass resonance.  
    
    In this paper we report an amplitude analysis and BF measurement of $D_{s}^{+} \rightarrow K^{+}K^{-}\pi^{+}$ (the inclusion of charge conjugates is implied) using a 3.19~fb$^{-1}$ data sample collected with the BESIII detector at a center-of-mass energy ($E_{{\rm CMS}}$) of 4.178 GeV.
    At this energy, the cross section for the $D_{s}^{*\pm}D_{s}^{\mp}$ final state in $e^{+}e^{-}$ annihilations is one order magnitude larger than that for $D_{s}^{+}D_{s}^{-}$~\cite{DsStrDs}.
    Moreover, the $D_{s}^{*\pm}$ decays are dominated by the process $D_{s}^{*\pm} \rightarrow \gamma D_{s}^{\pm}$~\cite{PDG}.
    Thus, the process $e^{+}e^{-} \rightarrow D_{s}^{*\pm}D_{s}^{\mp} \rightarrow D_{s}^{+}\gamma D_{s}^{-}$ is the main signal process.  
    Using a tagging technique~\cite{double-tag} (described in Sec.~\ref{DT-AA}), we get a nearly background-free data sample to use for an amplitude analysis and BF measurement.  
    The process $e^{+}e^{-} \rightarrow D_{s}^{+}D_{s}^{-}$ 
also contributes to the BF measurement.  
For the MIPWA (Sec.~\ref{MIPWA}), only the signal decay is reconstructed,
while for the amplitude analysis (Sec.~\ref{AA}) and BF measurement (Sec.~\ref{BF}) 
both the signal $D_{s}$ and the other $D_{s}$ are reconstructed.

    \section{BESIII Detector and Data Sets}
    \label{sec:detector_dataset}
    The BESIII detector is a magnetic spectrometer~\cite{BESIII} located at the Beijing Electron Positron Collider (BEPCII)~\cite{BEPCII}.
    The inner subdetectors are surrounded by a superconducting solenoidal magnet which provide a 1.0 T magnetic field.  Starting from the interaction point these consist of a main drift chamber (MDC), a plastic scintillator time-of-flight (TOF) system, a CsI(Tl) electromagnetic calorimeter (EMC).
    Charged particle identification is performed by combining the ionization energy loss ($dE/dx$) measured by the MDC and the time-of-flight measured by the TOF.  The EMC provides shower information to reconstruct photons.
Outside the solenoidal magnet is a multi-gap resistive-plate chamber system, which provides muon identification.  

Monte Carlo (MC) samples are produced with {\sc geant4}-based~\cite{GEANT4} software.
To assess background processes and determine detection efficiencies, we
produce and analyzed an inclusive MC sample, with size that is 40 times the integrated luminosity of data.
The sample includes all known open charm production processes, the continuum processes ( $e^{+}e^{-} \rightarrow q\bar{q}$, $q = u, d$ and $s$), Bhabha scattering, $\mu^{+}\mu^{-}$, $\tau^{+}\tau^{-}$, diphoton process and production of the $c\bar{c}$ resonances $J/\psi$, $\psi(3686)$ and $\psi(3770)$ via initial state radiation (ISR).
The generator {\sc conexc}~\cite{CONEXC} is used to model the open charm processes directly produced via $e^{+}e^{-}$ annihilation.
The simulation of ISR production of $\psi(3770)$, $\psi(3686)$ and $J/\psi$ is performed with {\sc kkmc}~\cite{KKMC}.
The known decays with BFs taken from the Particle Data Group (PDG)~\cite{PDG} are simulated with {\sc evtgen}~\cite{EVTGEN} and
the unknown decays are generated with the {\sc lundcharm} model~\cite{LUNDCHARM}.
Final-state radiation from charged tracks is produced by {\sc photos}~\cite{PHOTOS}.
Additionally, we generate two MC samples with $e^{+}e^{-} \rightarrow D_{s}^{(*)}D_{s}$, in which the $D_{s}^{+}$ meson decays into $K^{+}K^{-}\pi^{+}$ while the $D_{s}^{-}$ meson decays to one of the tag modes listed in
Table~\ref{ST-mass-window}, with size 600 times larger than the expected number of signal events in data.
The one with a uniform distribution of $D_{s}^{+} \rightarrow K^{+}K^{-}\pi^{+}$ decays over the phase space (PHSP) is called ``PHSP MC".
In the second sample, called ``signal MC", the $D_{s}^{+} \rightarrow K^{+}K^{-}\pi^{+}$ decay is generated according to the model obtained from the amplitude analysis presented in this paper.
PHSP MC is used to calculate the MC integrations, while signal MC is used to
check the fit performance, calculate the goodness of fit and estimate the
detection efficiency.
\section{Event Selection}
\label{chap:event_selection}
The polar angles ($\theta$) of charged tracks with respect to the beam axis must satisfy $|\!\cos\theta| < 0.93$.
Except for tracks from $K_{S}^{0}$ decays, the distances of closest approach to the beamspot for charged tracks in the transverse plane and along the beam direction must be less than 1~cm and 10~cm, respectively.

Photons are reconstructed from showers in the EMC.
The deposited energies of the photons from the endcap ($0.86 < |\!\cos\theta| < 0.92$) should be larger than 50~MeV and those of the photons from the barrel ($|\!\cos\theta| < 0.80$) should be larger than 25~MeV.
Furthermore, the shower should be detected within 700~ns after a beam crossing.

Candidates for $\pi^{0} (\eta)$ decay are reconstructed through $\pi^{0} \rightarrow \gamma\gamma$ ($\eta \rightarrow \gamma\gamma$).
The diphoton invariant mass $M_{\gamma\gamma}$ for $\pi^{0}$ ($\eta$) should be in the range of 0.115 $< M_{\gamma\gamma} <$ 0.150~GeV/$c^{2}$ (0.490 $< M_{\gamma\gamma} <$ 0.580~GeV/$c^{2}$).  
A kinematic fit constraining $M_{\gamma\gamma}$ to the $\pi^{0}$ or $\eta$ nominal mass~\cite{PDG} is performed,
and the $\chi^{2}$ of the corresponding fit should be less than 30 for $\pi^{0}$ or $\eta$ candidates.

Kaons and pions are identified by combining the $dE/dx$ information in the MDC and the time-of-flight from the TOF.
If the probability of the kaon hypothesis is larger than that of the pion hypothesis, the track is identified as a kaon.
Otherwise, the track is identified as a pion.
Any $\pi^{\pm}$ and $\pi^{0}$ candidates with momentum less than 0.1~GeV/$c$  are vetoed to remove soft $\pi^{\pm}$ and $\pi^{0}$ from $D^{*}$ decays.

Pairs of $\pi^{+}\pi^{-}$ are used to reconstruct $K_{S}^{0}$ mesons.
The polar angles $\theta$ of the two pions should satisfy $|\!\cos\theta| < 0.93$.
The distances of closest approach to the beamspot along the beam direction should be less than 20~cm. 
The invariant mass $m(\pi^{+}\pi^{-})$ of $\pi^{+}\pi^{-}$ pairs  should satisfy $0.487 < m(\pi^{+}\pi^{-}) < 0.511$ GeV/$c^{2}$.
A secondary vertex fit, constraining the pion candidate pair to a common 
vertex is performed to determine the decay length $L$ of the $K_{S}$.
We require $L/\sigma_{L} > $ 2, where $\sigma_{L}$ is the uncertainty on $L$. 

The $\eta^{\prime}$ candidates are reconstructed via the process $\eta^{\prime} \rightarrow \pi^{+}\pi^{-}\eta$.
Candidates with a $\pi^{+}\pi^{-}\eta$ invariant mass in the range of $[0.938, 0.978]$ GeV/$c^{2}$ are retained.

Tagged $D_{s}$ candidates are reconstructed from various combinations of $K^{\pm}$, $\pi^{\pm}$, $\eta$, $\eta^{\prime}$, $K_{S}^{0}$ and $\pi^{0}$, while the signal $D_{s}^+$ candidates are reconstructed from $K^{+}K^{-}\pi^{+}$ combinations.  
Candidates with an invariant mass in the mass window $[1.87, 2.06]$ GeV/$c^{2}$ and a recoiling mass $M_{{\rm rec}}$ in the mass window $[2.051, 2.180]$ GeV/$c^{2}$ are retained.
The recoiling mass $M_{{\rm rec}}$ is defined as:
\begin{linenomath}
\begin{equation}
    \begin{aligned}
    M_{\rm rec} &=  \bigg[\left(E_{{\rm CMS}}c^{-2} - \sqrt{\mid \vec p_{D_{s}}c^{-1} \mid^{2} + m_{D_{s}}^{2}}\,\right)^{2}
    \\
    &- \mid\vec p_{D_{s}}c^{-1} \!\mid^{2}\bigg]^{\frac{1}{2}} \; , \label{con:inventoryflow}
    \end{aligned}
\end{equation}
\end{linenomath}
where $\vec p_{D_{s}}$ is the momentum of $D_{s}$ candidate in $e^{+}e^{-}$  center-of-mass system, $m_{D_{s}}$ is $D_{s}$ mass quoted from PDG~\cite{PDG}.
The requirement on $M_{\rm rec}$ is chosen to retain both 
the monochromatic $D_s$ that are produced directly from the $e^+e^-$ 
collision as well as the broader distribution that arises from 
$D_s^{*\pm} \to D_s^{\pm} \gamma$ decays.  

\section{Model-independent Partial Wave Analysis in the Low $K^{+}K^{-}$ Mass Region}
\label{MIPWA}
A MIPWA is performed to determine the S-wave lineshape near the threshold of $K^{+}K^{-}$ mass spectrum.
As the background contamination is rather low in this region of the Dalitz plot, 
and higher statistics are needed in this MIPWA, the event selection in this section is different from those
in the amplitude analysis (Sec.~\ref{AA}) and BF measurement (Sec.~\ref{BF}).

In the data sample used in the MIPWA, $D_{s}^{+} \rightarrow K^{+}K^{-}\pi^{+}$ candidates are reconstructed according to the selections in Sec.~\ref{chap:event_selection}.
The daughter tracks are further subjected to a 1C kinematic fit constraining them to the nominal $D_{s}^{+}$ mass from PDG~\cite{PDG}; 
selection of the best $D_{s}^{+} \rightarrow K^{+}K^{-}\pi^{+}$ candidate is based on the smallest $\chi^{2}$ in cases of multiple candidates.
The best photon candidate for the decay $D_{s}^{*\pm} \rightarrow D_{s}^{\pm}\gamma$, 
is obtained via the recoiling mass against the signal $D_{s}$ and the photon: 
\begin{linenomath}
\begin{equation}
\begin{aligned}
   M_{{\rm oth}} &= \bigg[ \left( E_{\rm CMS}c^{-2} - \sqrt{| \vec p_{D_{s}}c^{-1} |^{2} + m_{D_{s}}^{2}} -
   E_{\gamma}c^{-2}\right)^{2} \\ 
   &- | \vec p_{D_{s}}c^{-1} + \vec p_{\gamma}c^{-1}|^{2} \bigg]^{\frac{1}{2}} \; , \label{con:inventoryflow1}
\end{aligned}
\end{equation}
\end{linenomath}
where $E_{\gamma}$ and $\vec p_{\gamma}$ refer to the energy and momentum of a certain photon candidate in $e^{+}e^{-}$ center-of-mass system, respectively.  
The photon candidate resulting in the $M_{{\rm oth}}$ closest to the nominal $D_{s}$ mass is chosen as the best one.  

\begin{figure}[htbp]
 \centering
 \mbox{
  \begin{overpic}[width=0.48\textwidth]{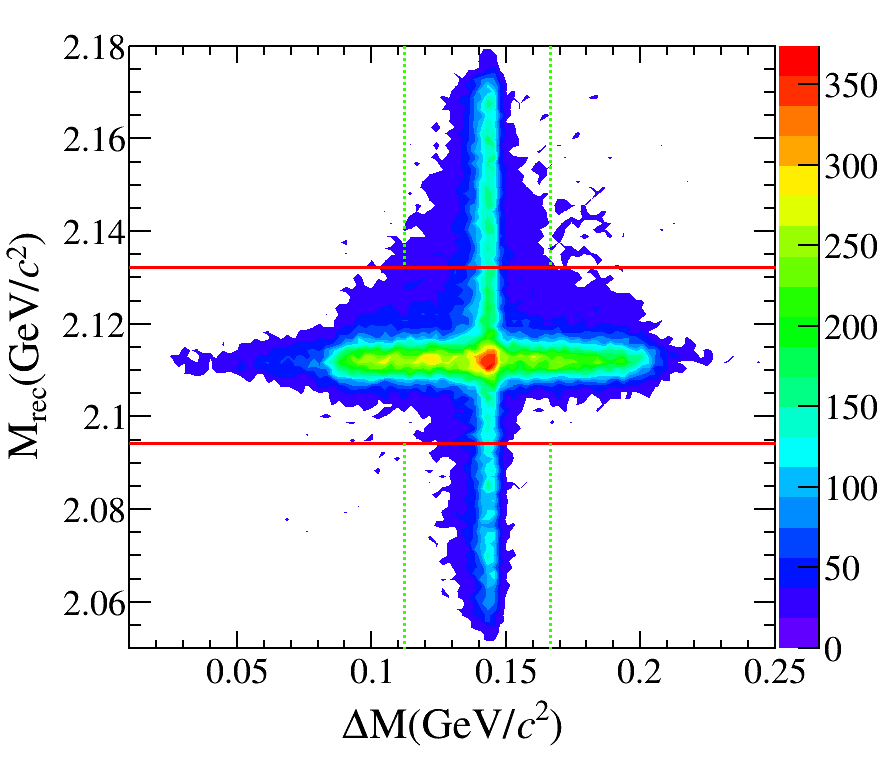}
  \end{overpic}
 }
 \caption{Two dimensional plane of ${M_{{\rm rec}}}$ versus $\Delta{M} \equiv M(D_{s}^{+}\gamma) - m_{{\rm sig}}$ from the simulated ${D_{s}^{+} \rightarrow K^{+}K^{-}\pi^{+}}$ decays. 
 The red solid (green dashed) lines mark the mass window for the $D_{s}^{+}$ category 1 (category 2) around the $M_{{\rm rec}}$ ($\Delta{M}$) peak.} 
\label{2DAll}
\end{figure}

A multi-variate analysis (MVA) method is used to suppress background from the $q\bar{q}$ continuum and other open charm processes. 
With the gradient boosted decision tree classifier (BDTG) provided by TMVA~\cite{TMVA}, we train the MVA separately with two sets of variables for the two categories depending on the $D_{s}^{+}$ origin.
Two categories of events are selected in an $M_{ {\rm rec}}$ versus $\Delta M$ 2D plane, where $\Delta M \equiv M(D_{s}^{+}\gamma) - m_{{\rm sig}}$, $m_{{\rm sig}}$ is the invariant mass of signal $D_{s}$ and $M(D_{s}^{+}\gamma)$ refers to the invariant mass of $D_{s}^{+}$ and the photon from $D_{s}^{*+} \rightarrow D_{s}^{+}\gamma$, as shown in Fig.~\ref{2DAll}.
The events that satisfy $|M_{{\rm rec}} - 2.112| <$ 0.02 GeV/$c^{2}$ (the region within the red solid lines in Fig.~\ref{2DAll}) are denoted as category 1, 
while the events that satisfy $|M_{{\rm rec}} - 2.112|>$ 0.02 GeV/$c^{2}$ and 0.112 $<\Delta M <$ 0.167 GeV/$c^{2}$ (the region within the green dashed lines in Fig.~\ref{2DAll}) are denoted as category 2.

\begin{figure}[htbp]
 \centering
 \mbox{
  \begin{overpic}[width=0.48\textwidth]{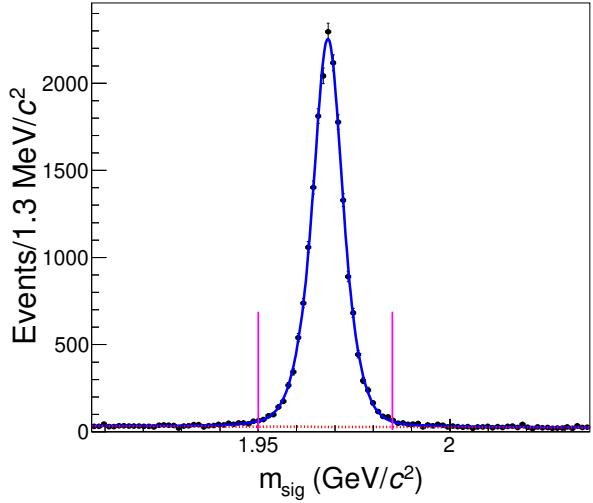}
  \end{overpic}
 }
 \caption{The fit to the signal $D_{s}$ invariant mass $m_{sig}$ (the dots with error bars) after BDTG requirement. The area between the pink lines is the signal area of the sample for MIPWA.
     Here, $m_{sig}$ is the mass without 1C kinematic fit correction.
     The signal shape is the MC-simulated shape convolved with a double Gaussian function and the background shape (red dotted line) is second-order Chebychev polynomial.
 }
\label{MIPWA-ST}
\end{figure}

For category 1, the BDTG takes three discriminating variables as input: the recoiling mass $M_{{\rm rec}}$, the total momentum of the un-reconstructed objects in the event (not part of the $D_{s}^{+} \rightarrow K^{+}K^{-}\pi^{+}$ candidate) and the energy of the photon from $D_{s}^{*}$.
For category 2, the BDTG takes three additional variables as input: $\Delta M$, $M_{{\rm rec}}^{\prime}$ and the total number of charged tracks and neutrals in an event $N_{{\rm tracks}}$.
Here, $M_{{\rm rec}}^{\prime}$ is defined as $M_{{\rm rec}}^{\prime} =
\left[ { {\left(E_{{\rm CMS}}c^{-2}  - \sqrt{|\vec p_{D_{s}\gamma}c^{-1}|^{2} + m_{D_{s}^{*}}^{2}} \; \right) }^{2} - | \vec p_{D_{s}\gamma}c^{-1}|^{2}}\right]^{\frac{1}{2}}$, where $\vec p_{D_{s}\gamma}$ is the momentum of the $D_{s}\gamma$ combination 
in $e^{+}e^{-}$ center-of-mass system
and $m_{D_{s}^{*}}$ is the nominal ${D_{s}^{*}}$ mass.
According to studies with the inclusive MC sample, the BDTG requirement gives a relatively pure sample (background less than 4\%) and the background ratios of category 1 and category 2 are similar. 
After applying the BDTG requirement, we fit to the candidate signal $D_{s}$ invariant mass for both category 1 and category 2 events.  
The signal shape is modeled with the MC-simulated shape convolved with a double Gaussian function to account for the difference between data and MC simulation, while the background is described with a second-order Chebychev polynomial.  
This fit gives a background yield in signal region ($1.950 < m_{{\rm sig}} < 1.986$ GeV/$c^{2}$) of 766 $\pm$ 30 and a corresponding signal yield of 18600 $\pm$ 141, as shown in Fig.~\ref{MIPWA-ST}.  

Assuming $N$ is the number of events for a given mass interval of $m(K^{+}K^{-})$, the angular distribution $\frac{dN}{d\cos\Theta}$ can be expanded in terms of spherical harmonic functions:
\begin{linenomath}
    \begin{equation}
        \frac{dN}{d\cos\Theta} = 2\pi\sum\limits_{k=0}^{L_{{\rm max}}}\left\langle Y_{k}^{0}\right\rangle Y_{k}^{0}(\cos\Theta),\label{expansion}
    \end{equation}
\end{linenomath}
where $L_{{\rm max}} = 2 \ell_{{\rm max}}$, and $\ell_{{\rm max}}$ is the maximum orbital angular momentum quantum number required to describe the $K^{+}K^{-}$ system at $m({K^{+}K^{-}})$ (e.g. $\ell_{{\rm max}}$ =1 when only S-, P-wave are considered), $\Theta$ is the angle between the $K^{+}$ direction in the $K^{+}K^{-}$ rest frame and the prior direction of the $K^{+}K^{-}$ system in the $D_{s}^{+}$ rest frame, $Y_{k}^{0}(\cos\Theta) = \sqrt{(2k+1)/4\pi}P_{k}(\cos\Theta)$ are harmonic functions, $P_{k}(\cos\Theta)$ is $k$-th order Legendre polynomial.

    The background contribution is subtracted from the selected sample
using the shape of the $m(K^{-}\pi^{+})$ versus $m(K^{+}K^{-})$ distribution from the inclusive MC sample,
while the background normalization is fixed according to the fit results (see Fig.~\ref{MIPWA-ST}). 
    After that the distribution $dN/d\cos\Theta$ of data are corrected for efficiency and phase space.
    The distribution $m(K^{-}\pi^{+})$ versus $m(K^{+}K^{-})$ of PHSP MC is used to calculate the efficiency.
    The correction for the Lorentz invariant phase space factor is calculated as $\rho_{KK} = \sqrt{1 - 4m^{2}_{K}/m^{2}(K^{+}K^{-})}$, 
    where $m_{K}$ is the nominal mass of $K^{\pm}$~\cite{PDG}.
    The harmonic functions $Y_{k}^{0}(\cos\Theta)$ are normalized as follows:
\begin{linenomath}
    \begin{equation}
        \int_{-1}^{1}Y_{k}^{0}(\cos\Theta)Y_{j}^{0}(\cos\Theta) d\cos\Theta  = \frac{\delta_{kj}}{2\pi}.\label{sh-normalizations}
    \end{equation}
\end{linenomath}
Considering the orthogonality condition, we can obtain the expansion coefficients according to Eqs.~(\ref{expansion}) and~(\ref{sh-normalizations}):
\begin{linenomath}
    \begin{equation}
        \left\langle Y_{k}^{0} \right\rangle = \int_{-1}^{1}Y_{k}^{0}(\cos\Theta) \frac{dN}{d\cos\Theta} d\cos\Theta. \label{expansion-coefficients}
    \end{equation}
\end{linenomath}
    In this section, the formalism $\sum\limits_{n=1}^{N}Y_{k}^{0}(\cos\Theta_{n})$ is used to calculate the integral, where $\Theta_{n}$ refers to the $\Theta$ of the $n$-th event.

\begin{figure*}[htbp]
        \centering
        \mbox{
            \begin{overpic}[width=0.98\textwidth]{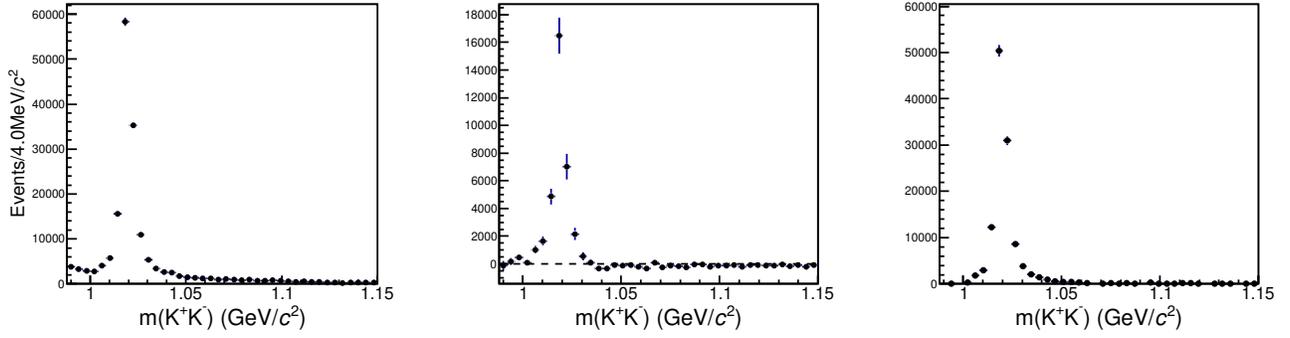}
            \end{overpic}
        }
        \caption{ The distribution of (a) $\left\langle Y_{0}^{0} \right\rangle$, (b) $\left\langle Y_{1}^{0} \right\rangle$ and (c) $\left\langle Y_{2}^{0} \right\rangle$ in $K^{+}K^{-}$ threshold region.}
        \label{Y0}
    \end{figure*}

    According to $\left\langle Y_{k}^{0} \right\rangle = \sum\limits_{n=1}^{N}Y_{k}^{0}(\cos\Theta_{n})$, one obtains the distribution of $\left\langle Y_{k}^{0} \right\rangle$ for $k$ = 0, 1 and 2 at the low end of $K^{+}K^{-}$ mass spectrum (0.988 $< m(K^{+}K^{-}) < $ 1.15 GeV/$c^{2}$), as shown in Fig.~\ref{Y0}.

\begin{figure*}[htbp]
        \centering
        \includegraphics[width=0.45\textwidth]{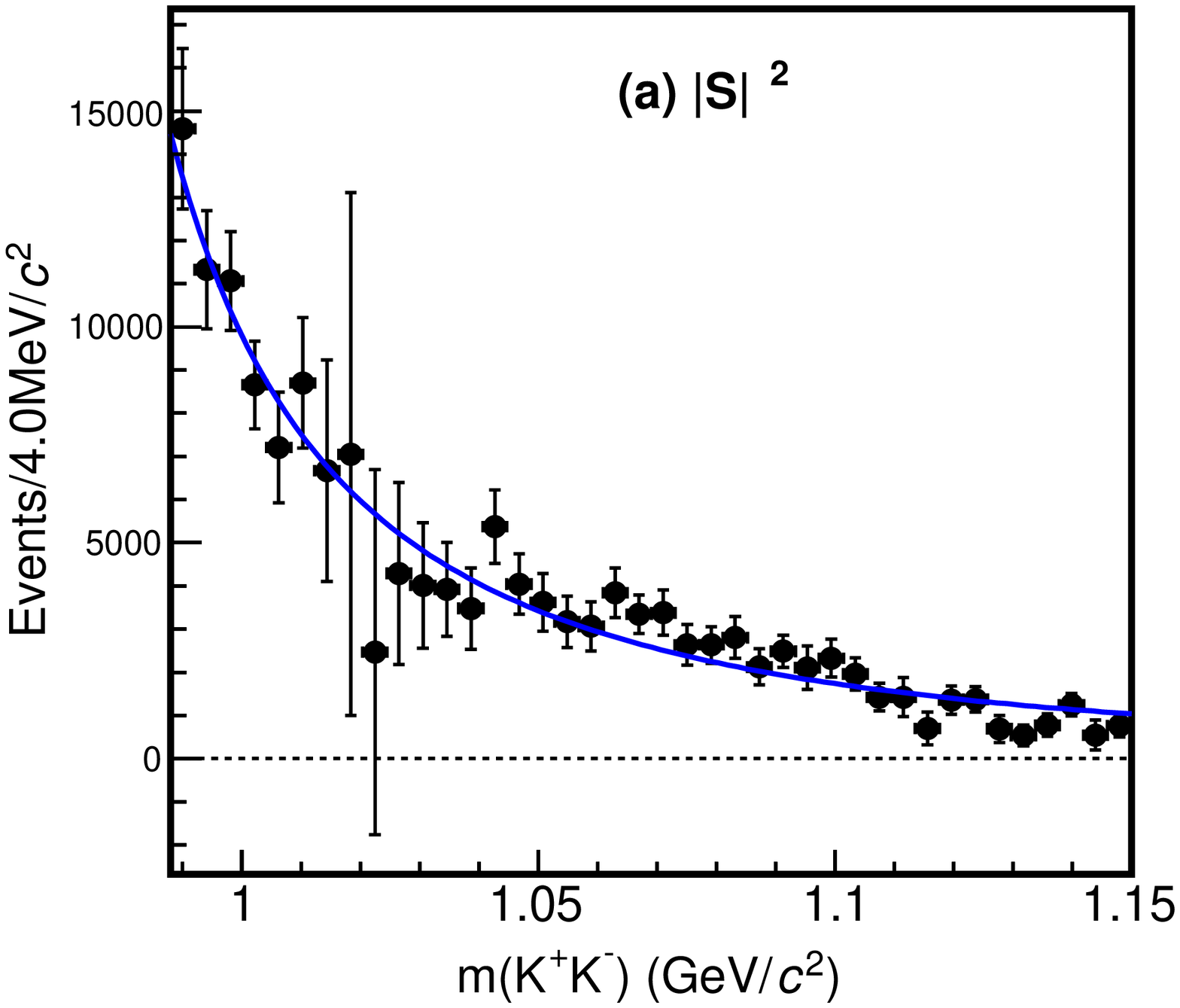}
        \includegraphics[width=0.45\textwidth]{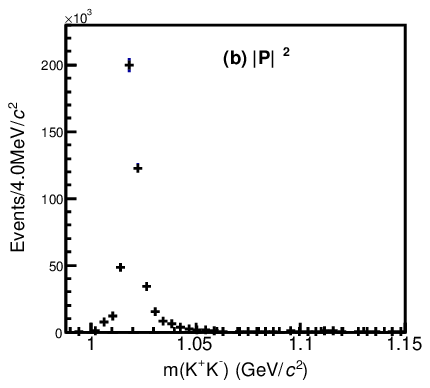}
        \includegraphics[width=0.45\textwidth]{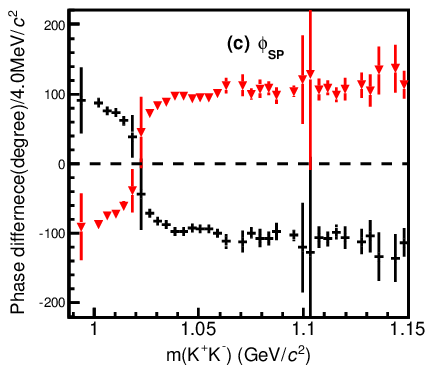}
        \includegraphics[width=0.45\textwidth]{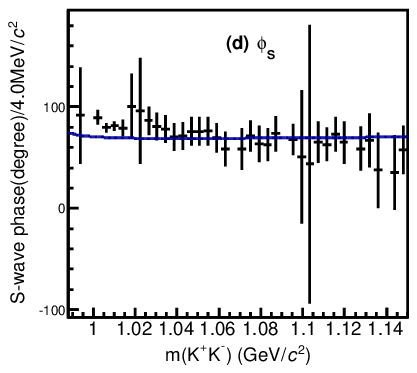}
        \caption{The distribution of (a) $\left|{\rm S}\right|^{2}$, (b) $\left|{\rm P}\right|^{2}$, (c) $\phi_{{\rm SP}}$ and (d) $\phi_{{\rm S}}$ in the threshold region of $m(K^{+}K^{-})$.
            The description of the fit in (a) can be found in the text. 
            The solid line in (d) shows the phase of $S(980)$ amplitude obtained from the amplitude analysis (Sec.~\ref{AA}).
        }
        \label{SP}
    \end{figure*}

    Assuming that only S- and P-wave amplitudes are necessary at the low end of $K^{+}K^{-}$ mass spectrum, the distribution  $dN/d\cos\Theta$ can also be written in terms of the partial wave amplitudes:
    \begin{linenomath}
        \begin{equation}
            \frac{dN}{d\cos\Theta} = 2\pi\left|{\rm S}Y_{0}^{0}(\cos\Theta) + {\rm P}Y_{1}^{0}(\cos\Theta)\right|^{2},\label{SP-distribution}
        \end{equation}
    \end{linenomath}
    where S and P refer to the amplitudes of S-wave and P-wave, respectively.
    Comparing Eqs.~(\ref{expansion}) and ~(\ref{SP-distribution})~\cite{PRD56-7299}, we can obtain 
    \begin{linenomath}
        \begin{equation}
            \begin{array}{lr}
                \left|{\rm S}\right|^{2} = \sqrt{4\pi}\left\langle Y_{0}^{0}\right\rangle - \sqrt{5\pi}\left\langle Y_{2}^{0}\right\rangle, &\\
                                                                                                                                      &\\
                \cos\phi_{{{\rm SP}}} = \frac{\left\langle Y_{1}^{0}\right\rangle}{ \sqrt{ \left(2\left\langle Y_{0}^{0}\right\rangle - \sqrt{5}\left\langle Y_{2}^{0}\right\rangle\right)\sqrt{5}\left\langle Y_{2}^{0}\right\rangle  }}   , &\\
                                                                                                                                                                                                                                              &\\
                \left|{\rm P}\right|^{2} = \sqrt{5\pi}\left\langle Y_{2}^{0}\right\rangle, &
            \end{array}\label{SP-RES} 
        \end{equation}
    \end{linenomath}
    where $\phi_{{\rm SP}} = \phi_{{\rm S}} - \phi_{{\rm P}}$ is the phase difference between S-wave and P-wave, $\phi_{{\rm S}}$ and $\phi_{{\rm P}}$ are the phases of S-wave and P-wave, respectively.
    Calculating $\left|{\rm S}\right|^{2}$, $\phi_{{\rm SP}}$ and $\left|{\rm P}\right|^{2}$ in every mass interval of $m(K^{+}K^{-})$ in the threshold region,
    the distribution of $\left|{\rm S}\right|^{2}$, $\left|{\rm P}\right|^{2}$, $\phi_{{\rm SP}}$ and $\phi_{{\rm S}}$ can be obtained,  as shown in Fig.~\ref{SP}.
    There are two curves in Fig.~\ref{SP}(c) because of the sign ambiguity of $\phi_{{\rm SP}}$ extracted from $\cos\phi_{{\rm SP}}$.

    We found that the Flatt\' e parameterization~\cite{Flatte} is insensitive to the $\pi\pi$ or $\pi\eta$ coupling or the coupling induced between them while fitting the distribution of $\left|{\rm S}\right|^{2}$.
    Therefore, the lineshape of $S(980)$ is empirically parameterized with the following formula:
    \begin{linenomath}
    \begin{equation}
        A_{S(980)} = \frac{1}{m_{0}^{2} - m^{2} -im_{0}\Gamma_{0}\rho_{KK}}. \label{S980-RBW}
    \end{equation}
    \end{linenomath}
    Fitting the distribution of $\left|{\rm S}\right|^{2}$ in Fig.~\ref{SP}(a) with $|A_{S(980)}|^{2}$, we can obtain the values of $m_{0}$ and $\Gamma_{0}$:
    \begin{linenomath}
    \begin{equation}
        \begin{aligned}
            m_{0} &= (0.919 \pm 0.006_{{\rm stat}}) \ {\rm GeV}/c^{2}, \\
            \Gamma_{0} &= (0.272 \pm 0.040_{{\rm stat}}) \ {\rm GeV}. 
        \end{aligned}\label{S-wave parameters} 
    \end{equation}
    \end{linenomath}
    Figure~\ref{SP}(a) shows the fit result. The $\chi^{2}/{\rm NDF}$ of the fit is 44.46/38 = 1.17.

The $S(980)$ mass central value obtained from the fit
is much lower than the threshold of $m(K^{+}K^{-})$ (about 0.988 GeV/$c^{2}$).
Therefore the distribution of $\phi_{{\rm S}}$ is expected to be roughly constant.
The phase $\phi_{{\rm P}}$ of the $\phi(1020)$ is given by Eq.~(\ref{RBW}) 
in Sec.~\ref{likelihood}; it increases rapidly near the $\phi(1020)$ peak 
because of its narrow width.  
Then the sign ambiguity of $\phi_{{\rm SP}}$ is solved by choosing the black curve
in Fig.~\ref{SP}(c), which decreases rapidly near the  mass 
of the $\phi(1020)$, ensuring that 
$\phi_{{\rm S}} = \phi_{{\rm P}} + \phi_{{\rm SP}}$ is roughly constant.  
The resulting phase of the $S(980)$, $\phi_{{\rm S}}$, is shown in Fig.~\ref{SP}(d). 
The solid line in Fig.~\ref{SP}(d) shows the phase of $S(980)$ amplitude obtained from the amplitude analysis (described in Sec.~\ref{AA}).
We can see that the shapes of $\phi_{{\rm S}}$ from model-independent analysis and amplitude analysis are consistent. 
The values of $\left|{\rm S}\right|^{2}$ (arbitrary units), $\left|{\rm P}\right|^{2}$ (arbitrary units) and $\phi_{{\rm S}}$ in every mass interval of the threshold region are listed in Table~\ref{SP-values}.
    \begin{table*}[htbp]
        \caption{
            The values of $\left|{\rm S}\right|^{2}$ (arbitrary units), $\left|{\rm P}\right|^{2}$ (arbitrary units) and $\phi_{{\rm S}}$; the units chosen preserve the relative $\left|{\rm S}\right|^{2}$ and $\left|{\rm P}\right|^{2}$ sizes.  
            Uncertainties in the table are statistical only.
            Some values of $\phi_{{\rm S}}$ are not listed in the table because the values of $\left\langle Y_{2}^{0}\right\rangle$ in the corresponding mass intervals are negative and a physical solution for $\phi_{{\rm SP}}$ cannot be found according to Eq.~(\ref{SP-RES}). 
        }
        \renewcommand\arraystretch{0.9}
        \label{SP-values}
        \begin{center}
            \begin{tabular}{c|ccc}
                \hline\hline
                $m(K^{+}K^{-})$ (GeV/$c^{2}$) & $\left|{\rm S}\right|^{2}$ (arbitrary units) & $\left|{\rm P}\right|^{2}$ (arbitrary units) & $\phi_{{\rm S}}$ (degrees)\\
                \hline
                $[0.988,\ 0.992]$   &	14593$\pm$1860&	                        \hphantom{0}$-$1137$\pm$1401&	            - \\ 	
                $[0.992,\ 0.996]$   &	11326$\pm$1364&	                        \hphantom{000}168$\pm$1027&	            \hphantom{0}92$\pm$\hphantom{0}48 \\ 	
                $[0.996,\ 1.000]$   &	11064$\pm$1143&	                        \hphantom{00}$-$531$\pm$\hphantom{0}850&	 - \\ 	
                $[1.000,\ 1.004]$   &	\hphantom{0}8659$\pm$1015&	            \hphantom{00}1006$\pm$\hphantom{0}748&	\hphantom{0}90$\pm$\hphantom{00}7 \\ 	
                $[1.004,\ 1.008]$   &	\hphantom{0}7207$\pm$1281&	            \hphantom{00}7292$\pm$1003&	            \hphantom{0}80$\pm$\hphantom{00}5 \\ 	
                $[1.008,\ 1.012]$   &	\hphantom{0}8703$\pm$1509&	            \hphantom{0}11746$\pm$1200&	            \hphantom{0}81$\pm$\hphantom{00}5 \\ 	
                $[1.012,\ 1.016]$   &	\hphantom{0}6669$\pm$2565&	            \hphantom{0}48763$\pm$2066&	            \hphantom{0}79$\pm$\hphantom{00}8 \\ 	
                $[1.016,\ 1.020]$   &	\hphantom{0}7051$\pm$6057&	            199740$\pm$5048	&                        101$\pm$\hphantom{0}32 \\ 	
                $[1.020,\ 1.024]$   &	\hphantom{0}2466$\pm$4232&	            122645$\pm$3520	&                        \hphantom{0}96$\pm$\hphantom{0}52 \\ 	
                $[1.024,\ 1.028]$   &	\hphantom{0}4292$\pm$2108&	            \hphantom{0}34363$\pm$1748&	            \hphantom{0}87$\pm$\hphantom{0}14 \\ 	
                $[1.028,\ 1.033]$   &	\hphantom{0}4009$\pm$1455&	            \hphantom{0}15046$\pm$1212&	            \hphantom{0}81$\pm$\hphantom{0}13 \\ 	
                $[1.033,\ 1.037]$   &	\hphantom{0}3922$\pm$1088&	            \hphantom{00}8108$\pm$\hphantom{0}887&	\hphantom{0}78$\pm$\hphantom{0}14 \\ 	
                $[1.037,\ 1.041]$   &	\hphantom{0}3480$\pm$\hphantom{0}944&	\hphantom{00}5945$\pm$\hphantom{0}768&	\hphantom{0}70$\pm$\hphantom{0}14 \\ 	
                $[1.041,\ 1.045]$   &	\hphantom{0}5376$\pm$\hphantom{0}854&	\hphantom{00}3707$\pm$\hphantom{0}678&	\hphantom{0}71$\pm$\hphantom{0}14 \\ 	
                $[1.045,\ 1.049]$   &	\hphantom{0}4043$\pm$\hphantom{0}696&	\hphantom{00}2103$\pm$\hphantom{0}551&	\hphantom{0}76$\pm$\hphantom{0}14 \\ 	
                $[1.049,\ 1.053]$   &	\hphantom{0}3621$\pm$\hphantom{0}665&	\hphantom{00}1858$\pm$\hphantom{0}530&	\hphantom{0}76$\pm$\hphantom{0}14 \\ 	
                $[1.053,\ 1.057]$   &	\hphantom{0}3167$\pm$\hphantom{0}599&	\hphantom{00}1680$\pm$\hphantom{0}467&	\hphantom{0}76$\pm$\hphantom{0}14 \\ 	
                $[1.057,\ 1.061]$   &	\hphantom{0}3063$\pm$\hphantom{0}569&	\hphantom{00}1333$\pm$\hphantom{0}448&	\hphantom{0}70$\pm$\hphantom{0}15 \\ 	
                $[1.061,\ 1.065]$   &	\hphantom{0}3841$\pm$\hphantom{0}582&	\hphantom{000}685$\pm$\hphantom{0}461&	\hphantom{0}59$\pm$\hphantom{0}17 \\ 	
                $[1.065,\ 1.069]$   &	\hphantom{0}3343$\pm$\hphantom{0}439&	\hphantom{000}$-$45$\pm$\hphantom{0}324&	 - \\ 	
                $[1.069,\ 1.073]$   &	\hphantom{0}3377$\pm$\hphantom{0}525&	\hphantom{000}395$\pm$\hphantom{0}413&	\hphantom{0}59$\pm$\hphantom{0}21 \\ 	
                $[1.073,\ 1.077]$   &	\hphantom{0}2635$\pm$\hphantom{0}474&	\hphantom{000}684$\pm$\hphantom{0}368&	\hphantom{0}71$\pm$\hphantom{0}15 \\ 	
                $[1.077,\ 1.081]$   &	\hphantom{0}2632$\pm$\hphantom{0}426&	\hphantom{000}357$\pm$\hphantom{0}320&	\hphantom{0}64$\pm$\hphantom{0}18 \\ 	
                $[1.081,\ 1.085]$   &	\hphantom{0}2802$\pm$\hphantom{0}485&	\hphantom{000}647$\pm$\hphantom{0}377&	\hphantom{0}63$\pm$\hphantom{0}16 \\ 	
                $[1.085,\ 1.089]$   &	\hphantom{0}2121$\pm$\hphantom{0}421&	\hphantom{000}287$\pm$\hphantom{0}332&	\hphantom{0}74$\pm$\hphantom{0}18 \\ 	
                $[1.089,\ 1.093]$   &	\hphantom{0}2487$\pm$\hphantom{0}369&	\hphantom{00}$-$185$\pm$\hphantom{0}278&	 - \\ 	
                $[1.093,\ 1.097]$   &	\hphantom{0}2105$\pm$\hphantom{0}505&	\hphantom{00}1041$\pm$\hphantom{0}409&	\hphantom{0}68$\pm$\hphantom{0}15 \\ 	
                $[1.097,\ 1.101]$   &	\hphantom{0}2326$\pm$\hphantom{0}440&	\hphantom{000}100$\pm$\hphantom{0}355&	\hphantom{0}51$\pm$\hphantom{0}66 \\ 	
                $[1.101,\ 1.105]$   &	\hphantom{0}1962$\pm$\hphantom{0}369&	\hphantom{0000}47$\pm$\hphantom{0}286&	\hphantom{0}44$\pm$137 \\ 	
                $[1.105,\ 1.109]$   &	\hphantom{0}1422$\pm$\hphantom{0}323&	\hphantom{000}216$\pm$\hphantom{0}246&	\hphantom{0}65$\pm$\hphantom{0}21 \\ 	
                $[1.109,\ 1.114]$   &	\hphantom{0}1420$\pm$\hphantom{0}453&	\hphantom{000}777$\pm$\hphantom{0}377&	\hphantom{0}63$\pm$\hphantom{0}17 \\ 	
                $[1.114,\ 1.118]$   &	\hphantom{00}697$\pm$\hphantom{0}377&	\hphantom{000}903$\pm$\hphantom{0}307&	\hphantom{0}73$\pm$\hphantom{0}17 \\ 	
                $[1.118,\ 1.122]$   &	\hphantom{0}1351$\pm$\hphantom{0}330&	\hphantom{000}234$\pm$\hphantom{0}257&	\hphantom{0}65$\pm$\hphantom{0}21 \\ 	
                $[1.122,\ 1.126]$   &	\hphantom{0}1373$\pm$\hphantom{0}297&	\hphantom{000}$-$60$\pm$\hphantom{0}229&	 - \\ 	
                $[1.126,\ 1.130]$   &	\hphantom{00}690$\pm$\hphantom{0}312&	\hphantom{000}340$\pm$\hphantom{0}255&	\hphantom{0}59$\pm$\hphantom{0}22 \\ 	
                $[1.130,\ 1.134]$   &	\hphantom{00}535$\pm$\hphantom{0}246&	\hphantom{000}130$\pm$\hphantom{0}197&	\hphantom{0}67$\pm$\hphantom{0}27 \\ 	
                $[1.134,\ 1.138]$   &	\hphantom{00}772$\pm$\hphantom{0}261&	\hphantom{000}205$\pm$\hphantom{0}199&	\hphantom{0}38$\pm$\hphantom{0}37 \\ 	
                $[1.138,\ 1.142]$   &	\hphantom{0}1246$\pm$\hphantom{0}266&	\hphantom{000}$-$71$\pm$\hphantom{0}200&	 - \\    
                $[1.142,\ 1.146]$   &	\hphantom{00}545$\pm$\hphantom{0}350&	\hphantom{000}456$\pm$\hphantom{0}298&	\hphantom{0}35$\pm$\hphantom{0}37 \\ 	
                $[1.146,\ 1.150]$   &	\hphantom{00}763$\pm$\hphantom{0}262&	\hphantom{000}206$\pm$\hphantom{0}205&	\hphantom{0}58$\pm$\hphantom{0}24 \\ 	
                \hline\hline
            \end{tabular}
        \end{center}
    \end{table*}

    Systematic uncertainties considered for the MIPWA include: 
    \begin{itemize}
        \item Data-MC agreement for the BDTG output. 
            A control sample is obtained with same event selection as that in Sec.~\ref{DT-AA} due to its high purity,
            but without the kinematic fit criteria.
            The efficiency of data and MC samples from the BDTG requirement is then considered,
            where the efficiency of data (MC) is defined as $e_{{\rm data}} = \frac{N_{d1}}{N_{d0}}$ ($e_{{\rm MC}} = \frac{N_{M1}}{N_{M0}}$), where $N_{d0}$ ($N_{M0}$) and $N_{d1}$ ($N_{M1}$) are the number of events before and after applying the BDTG requirement.
            We can now correct the data sample with $\frac{e_{{\rm data}}}{e_{{\rm MC}}}$.  
            We fit the corrected shape of the $S(980)$ and take the shift of $m_{0}$ and $\Gamma_{0}$ as the systematic uncertainty.

        \item Background subtraction. 
            We change the bin size, fit range and replace the background shape with a third-order Chebychev polynomial in the fit shown in Fig.~\ref{MIPWA-ST}.
            New fits are performed and we take the quadrature sum of the shifts as the uncertainty of the background fraction.
            Then we vary the background fraction, (3.9 $\pm$ 0.3)\%, within its uncertainty and take the shift of the $S(980)$ fit results as the systematic uncertainty related to the background fraction. 
            The background shape of inclusive MC sample is also replaced with that of sideband ($1.90 < m_{{\rm sig}} < 1.95$ GeV/$c^{2}$ and $1.986 < m_{{\rm sig}} < 2.03$ GeV/$c^{2}$) for data to perform a fit and the shift is taken as the systematic uncertainty related to the background shape.
            The quadrature sum of the shifts of $m_{0}$ and $\Gamma_{0}$ are 0.002 GeV/$c^{2}$ and 0.001 GeV, respectively.

        \item Particle identification (PID) and tracking efficiency difference between data and MC simulation. 
           The PID efficiencies are studied using control samples of $e^{+}e^{-} \rightarrow K^{+}K^{-}K^{+}K^{-}$, $K^{+}K^{-}\pi^{+}\pi^{-}$,  $K^{+}K^{-}\pi^{+}\pi^{-}\pi^{0}$, $\pi^{+}\pi^{-}\pi^{+}\pi^{-}$
           and $\pi^{+}\pi^{-}\pi^{+}\pi^{-}\pi^{0}$, 
           while a control sample of $e^{+}e^{-} \rightarrow K^{+}K^{-}\pi^{+}\pi^{-}$ is used for the study of tracking efficiencies. 
           In these control samples, all final particles are reconstructed with the selection criteria mentioned in
           Sec.~\ref{chap:event_selection} except the target kaon (pion). The total number of the target kaon (pion) are inferred by fitting the missing mass distributions
           while the number of the reconstructed target kaon (pion) is determined by applying corresponding selection criteria. 
           The efficiency difference of data and MC samples is assigned as the associated systematic uncertainty.
           These efficiencies are also used in the amplitude analysis (Sec.~\ref{AA})
           and BF measurement (Sec.~\ref{BF-sys}). 
           We weight each event with the data/MC efficiency differences and fit the shape of the $S(980)$.
           The shift of $m_{0}$ and $\Gamma_{0}$ are 0.001 GeV/$c^{2}$ and 0.013 GeV, respectively.

        \item The $f_{0}(1370)$ contribution. 
            The $f_{0}(1370)$ contribution in the $S(980)$ region was subtracted according to the measured FF.
            The shape of $f_{0}(1370)$ at the low end of $m_{K^{+}K^{-}}$ mass spectrum was obtained from the MC simulation.
            The interference effect was ignored.
            The resulting shifts of $m_{0}$ and $\Gamma_{0}$ are 0.001 GeV/$c^{2}$ and 0.003 GeV, respectively.

        \item Fit range. We vary the fit range from [0.988, 1.15] GeV/$c^{2}$ to [0.988, 1.145] GeV/$c^{2}$, which results in $m_{0}$ and $\Gamma_{0}$ shifts of 0.002 GeV/$c^{2}$ and 0.003 GeV, respectively.  
    
    \end{itemize}
    
    All of the systematic uncertainties mentioned above are summarized in Table~\ref{MIPWA-Sys}.
    The quadrature sum of the uncertainties is taken as the total uncertainty. 
    \begin{table*}[htbp]
        \caption{Systematic uncertainties of the partial wave analysis in the low $K^{+}K^{-}$ mass region.
        The quadrature sum of all contributions is taken as the total uncertainty.
        }
        \label{MIPWA-Sys}
        \begin{center}
            \begin{tabular}{l|ccc}
                \hline\hline
                Source   &                                                      $m_{0}$ (GeV/$c^{2}$)  &$\Gamma_{0}$ (GeV)\\
                \hline
                BDTG                     & 0.030                  &   0.020 \\
                Background subtraction   & 0.002                  &   0.001 \\
                PID and Tracking         & 0.001                  &   0.013 \\
                $f_{0}(1370)$            & 0.001                  &   0.003 \\
                Fit range                & 0.002                  &   0.003 \\

                \hline
                Total                                   & 0.030                  &   0.024\\
                \hline\hline
            \end{tabular}
        \end{center}
    \end{table*}
    We obtain the result for $m_{0}$ and $\Gamma_{0}$ with statistical and systematic errors to be:
    \begin{linenomath}
    \begin{equation}
        \begin{aligned}
            m_{0} &= (0.919 \pm 0.006_{{\rm stat}} \pm 0.030_{{\rm sys}}) \ {\rm GeV}/c^{2}, \\
            \Gamma_{0} &= (0.272 \pm 0.040_{{\rm stat}} \pm 0.024_{{\rm sys}})\ {\rm GeV},
        \end{aligned}\label{S-wave-sys} 
    \end{equation}
    \end{linenomath}
    which are consistent with the BaBar analysis~\cite{2011BABAR}.
    Note that $m_{0}$ and $\Gamma_{0}$ in Eq.~(\ref{S-wave-sys}) are only used for the parameterization of the $S(980)$ in Sec.~\ref{AA}.

\section{Amplitude Analysis}
\label{AA}

An unbinned maximum likelihood method is used to determine the intermediate resonance composition in the decay $D_{s}^{+} \rightarrow K^{+}K^{-}\pi^{+}$.
The likelihood function is constructed with a probability density function (PDF) which depends on the momenta of the three daughter particles. 

\subsection{Tag Technique in Amplitude Analysis}
\label{DT-AA}
As $D_{s}$ mesons are produced in pairs, $D_{s}$ mesons can be reconstructed with a tag technique 
which provides both single tag (ST) and double tag (DT) samples.
In the ST samples, only one $D_{s}^{-}$ meson is reconstructed through selected hadronic $D_{s}$ decays, the so-called tag modes.
The eight tag modes used in the amplitude analysis and BF measurement presented in Sec.~\ref{BF} are  
$D_{s}^{-} \rightarrow K^{+}K^{-}\pi^{-}$, $K_{S}^{0}K^{-}$, $ K_{S}^{0}K^{-}\pi^{+}\pi^{-}$, $K^{-}\pi^{+}\pi^{-}$, $K_{S}^{0}K^{+}\pi^{-}\pi^{-}$, $\pi^{+}\pi^{-}\pi^{-}$, $\pi^{-}\eta^{\prime}_{\pi^{+}\pi^{-}\eta_{\gamma\gamma}}$ and $K^{+}K^{-}\pi^{-}\pi^{0}$,
here, $\eta^{\prime}_{\pi^{+}\pi^{-}\eta_{\gamma\gamma}}$ denotes $\eta^{\prime} \rightarrow \pi^{+}\pi^{-}\eta$ with $\eta \rightarrow \gamma\gamma$.
In the DT samples, photons from the decay $D_{s}^{*\pm} \rightarrow D_{s}^{\pm}\gamma$, tag mode $D_{s}^{-}$ and signal $D_{s}^{+}$ 
(i.e., decays to $K^{+}K^{-}\pi^{+}$) are all fully reconstructed. 
A kinematic fit of $e^+ e^- \to D_s^{*\pm} D_s^{\mp} \to \gamma D_s^+D_s^-$
with $D_s^-$ decaying to one of the tag modes and $D_s^+$ decaying to the signal mode is performed.
We constrain the four-momentum of the $D_{s}^{*\pm}D_{s}^{\mp}$ system to the initial four-momentum of the electron-positron system and the invariant mass of the $D_{s}^{*\pm}$ to the corresponding PDG value~\cite{PDG}. 
This gives a total of five constraints (5C).
For each $D_s^+ D_s^- \gamma$ candidate,  the extra $\gamma$ is paired with both the tag and signal $D_s$ 
to form the $D_{s}^{*\pm}$, and the combination with the lower fit 
$\chi_{5{\rm C}}^{2}$ is retained as the presumably correct pairing.  
If there are multiple candidate $D_{s}^{*\pm}D_{s}^{\mp}$ pairs in an event, 
the candidate with minimum $\chi_{5{{\rm C}}}^{2}$ is selected as the best one.  
The invariant mass of signal $D_{s}$ ($m_{{\rm sig}}$) and tag $D_{s}$ ($m_{{\rm tag}}$) candidates are required to be within the mass regions 
shown in Table~\ref{ST-mass-window}.

\begin{table}[htbp]
    \caption{ The mass windows for the signal mode and various tag modes.}
    \label{ST-mass-window}
    \begin{center}
        \begin{tabular}{l|c}
            \hline\hline
            Mode & Mass window (GeV/$c^{2}$)  \\
            \hline
            $D_{s}^{-} \rightarrow K_{S}^{0}K^{-}$                          & [1.948, 1.991]    \\
            $D_{s}^{\pm} \rightarrow K^{\pm}K^{\mp}\pi^{\pm}$                       & [1.950, 1.986]    \\
            $D_{s}^{-} \rightarrow K^{+}K^{-}\pi^{-}\pi^{0}$                & [1.947, 1.982]    \\
            $D_{s}^{-} \rightarrow K_{S}^{0}K^{-}\pi^{+}\pi^{-}$            & [1.958, 1.980]    \\
            $D_{s}^{-} \rightarrow K_{S}^{0}K^{+}\pi^{-}\pi^{-}$            & [1.953, 1.983]    \\
            $D_{s}^{-} \rightarrow \pi^{-}\pi^{-}\pi^{+}$                   & [1.952, 1.984]    \\
            $D_{s}^{-} \rightarrow \pi^{-}\eta_{\pi^{+}\pi^{-}\eta_{\gamma\gamma}}^{\prime}$  & [1.940, 1.996]        \\
            $D_{s}^{-} \rightarrow K^{-}\pi^{+}\pi^{-}$                     & [1.953, 1.983]    \\
            \hline\hline
        \end{tabular}
    \end{center}
\end{table}

To ensure that all events fall within the physical region on the Dalitz plot, 
we perform a 7C fit where constraints on both signal and tag $D_s$ masses 
to the PDG values are added to the previous 5C constraints.  
The four-momenta of the tracks after 7C fit are used to perform 
the amplitude analysis.

\begin{figure}[htbp]
 \centering
 \mbox{
  \begin{overpic}[width=0.48\textwidth]{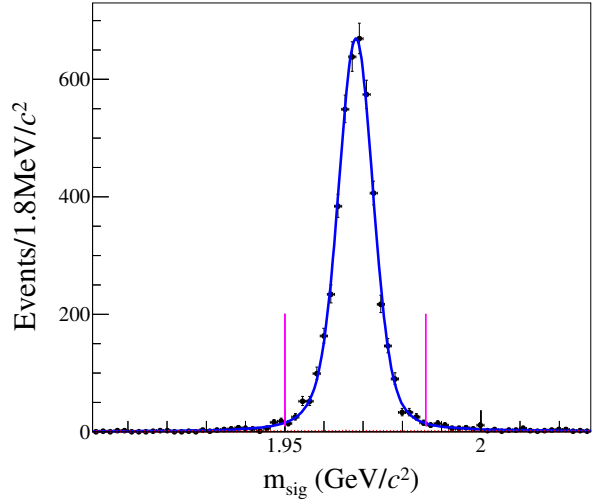}
  \end{overpic}
 }
 \caption{The fit to the signal $D_{s}$ invariant mass $m_{{\rm sig}}$ before the 7C kinematic fit (dots with error bars).  
          The area between the pink lines is the signal area of the sample for the amplitude analysis.}
\label{AA-mSig}
\end{figure}

\begin{figure}[htbp]
    \centering
    \mbox{
        \begin{overpic}[width=0.38\textwidth]{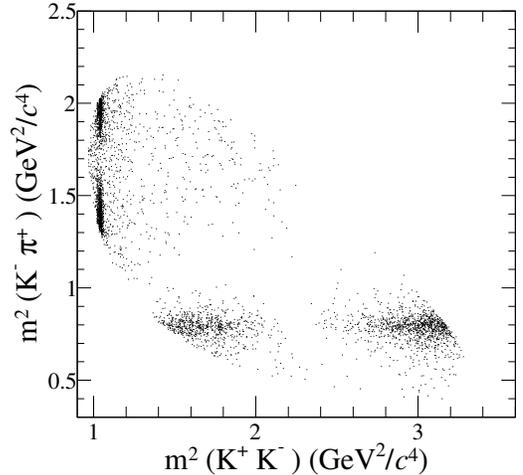}
        \end{overpic}
    }
    \caption{ The Dalitz plot of selected $D_{s}^{+} \rightarrow K^{+}K^{-}\pi^{+}$ candidates.}
    \label{dalitz}
\end{figure}

The background of the DT sample in the amplitude analysis is estimated using the inclusive MC sample.
The fit to the signal $D_{s}$ invariant mass without 7C kinematic fit gives the signal yield and purity, as shown in  Fig.~\ref{AA-mSig}.
In the fit, the signal shape is modeled with the MC-simulated shape convolved with a Gaussian function while the background is described with a second-order Chebychev polynomial.
There is no obvious peaking background in the signal region (1.950  $ < m_{{\rm sig}} < $1.986 GeV/$c^{2}$) and we obtain 4399 signal candidates with a purity of 99.6\%.  
Figure~\ref{dalitz} shows the Dalitz plot of the signal $D_{s}^{+} \rightarrow K^{+}K^{-}\pi^{+}$ candidates.

\subsection{Likelihood Function Construction}

\label{likelihood}

For a three-body process the amplitude $A_{n}(p)$ for the $n^{\rm{th}}$ 
mode may be written as 
    \begin{linenomath}
\begin{equation}
    A_{n}(p) = P_{n}(p)S_{n}(p)F_{n}^{r}(p)F_{n}^{D}(p), \label{base-amplitude}
\end{equation}
    \end{linenomath}
where $p$ refers to the set of the three daughter particles' four-momenta,
$P_{n}(p)$ is the propagator, $S_{n}(p)$ is the spin factor constructed with the covariant tensor formalism~\cite{covariant-tensors}, 
$F_{n}^{r}(p)$ and $F_{n}^{D}(p)$ are the Blatt-Weisskopf barrier factors for the intermediate resonance and $D_{s}$ meson decays, respectively.
According to the isobar formulation, the total amplitude $M(p,\alpha)$ is obtained by the coherent sum of all intermediate modes:
    \begin{linenomath}
\begin{equation}
M(p,\alpha)=\begin{matrix}\sum c_{n}A_{n}(p)\end{matrix}, \label{coherent-sum}
\end{equation}
    \end{linenomath}
where $\alpha$ is a set of fit parameters, which includes the $n^{\rm{th}}$ mode of complex coefficient $c_{n} = \rho_{n}e^{i\phi_{n}}$ ($\rho_{n}$ and $\phi_{n}$ are the magnitude and phase, respectively).
Then the signal PDF $f_{S}(p,\alpha)$ is given by: 
    \begin{linenomath}
\begin{equation}
    f_{S}(p,\alpha) = \frac{\epsilon(p)\left|M(p,\alpha)\right|^{2}R_{3}(p)}{\int \epsilon(p)\left|M(p,\alpha)\right|^{2}R_{3}(p)\,dp}, \label{signal-PDF}
\end{equation}
    \end{linenomath}
where $\epsilon(p)$ is the detection efficiency and $R_{3}(p)$ is the three-particle phase space density, which is defined as:
    \begin{linenomath}
\begin{equation}
    R_{3}(p)dp = (2\pi)^{4} \; \delta^{4}
    \left( p_{D_{s}} - \sum_{\beta=1}^{3}p_{\beta} \right)
    \prod_{\beta=1}^{3}\frac{d^{3}p_{\beta}}{ (2\pi)^{3}2E_{\beta}}, \label{three-body PHSP}
\end{equation}
    \end{linenomath}
where $\beta = 1, 2, 3$ is the index of the three daughter particles.
The likelihood function is given by
    \begin{linenomath}
\begin{equation}
    \begin{aligned}    
    \ln{\mathcal{L}}&= \begin{matrix}\sum\limits_{k}^{N_{{\rm data}}} \ln f_{S}(p^{k})\end{matrix}   \\
                    &= \sum\limits_{k}^{N_{{\rm data}}}\ln \left( \frac{\left|M(p,\alpha)\right|^{2}}{\int \epsilon(p)\left|M(p,\alpha)\right|^{2}R_{3}(p)\,dp}    \right)  \\
                    &+\sum\limits_{k}^{N_{{\rm data}}}\ln \left( R_{3}(p)\epsilon(p) \right), 
    \end{aligned}
    \label{likelihoodF}
\end{equation}
    \end{linenomath}
    where $N_{{\rm data}}$ is the number of candidate events in data.
    Note that the last term of Eq.~(\ref{likelihoodF}) is independent of the fit parameters and dropped during the log-likelihood fit.

    Background contribution is neglected in the amplitude analysis and the possible bias is included to the systematic uncertainties, see Sec.~\ref{AA-sys} below.
The normalization integral in Eq.~(\ref{likelihoodF}) is first determined by the following equation using PHSP MC events with a uniform distribution:
    \begin{linenomath}
\begin{equation}
    \begin{aligned}
        &\int \epsilon(p)\left|M(p,\alpha)\right|^{2}R_{3}(p)\,dp \\ 
        &\propto \frac{1}{N_{{{\rm MC,gen}}}} \begin{matrix}\sum\limits_{k_{{\rm MC}}}^{N_{{\rm MC,sel}}} \left|M(p^{k_{{\rm MC}}},\alpha)\right|^{2}\end{matrix},
    \end{aligned}
    \label{MC-intergral}
\end{equation}
    \end{linenomath}
    where $p^{k_{{\rm MC}}}$ is the $k_{{\rm MC}}^{\rm th}$ set of four-momenta.
    Here, $N_{{\rm MC,gen}}$ and $N_{{\rm MC,sel}}$ are the numbers of generated phase-space events and selected phase-space events, respectively.
A set of estimated fit parameters, denoted as $\alpha^{\prime}$, is obtained from a preliminary fit using the phase-space MC to evaluate the normalization integral.
The normalization integral is evaluated with signal MC samples:
    \begin{linenomath}
\begin{equation}
    \begin{aligned}
        &\int \epsilon(p)\left|M(p,\alpha)\right|^{2}R_{3}(p)\,dp \\ 
        &\propto \frac{1}{N_{{\rm MC}}} \begin{matrix}\sum\limits_{k_{{\rm MC}}}^{N_{{\rm MC\  }}} \frac{\left|M(p^{k_{{\rm MC}}},\alpha)\right|^{2}} { \left|M^{{\rm gen}}(p^{k_{{\rm MC}}},\alpha^\prime)\right|^{2} }\end{matrix}, 
    \end{aligned}
\label{sig-MC-intergral}
\end{equation}
    \end{linenomath}
    where $M^{{\rm gen}}(p^{k_{{\rm MC}}},\alpha^\prime)$ is the PDF modeled with $\alpha^{\prime}$ to generate signal MC and $N_{{\rm MC}}$ is the number of events in the MC sample.
The computational efficiency of  the MC integration is significantly
improved by evaluating the normalization integral with signal MC samples,
which intrinsically take into account the event selection acceptance and the detection resolution.
Correction factors $\gamma_{\epsilon}$ are introduced to correct for the bias caused by PID and tracking efficiency inconsistencies between data and MC simulation:
    \begin{linenomath}
\begin{equation}
    \gamma_{\epsilon} = \prod_{j} \frac{\epsilon_{j, data}(p)}{\epsilon_{j, MC}(p)}, \label{experimental-effect}
\end{equation}
    \end{linenomath}
where $j$ refers to PID or tracking, $\epsilon_{j, data}$ and $\epsilon_{j, MC}$ refer to the PID or tracking efficiencies for data and MC, respectively.
Taking the correction factors $\gamma_{\epsilon}$ into account, the normalization integral can be obtained by:

    \begin{linenomath}
\begin{equation}
    \begin{aligned}
        &\int \epsilon(p)\left|M(p,\alpha)\right|^{2}R_{3}(p)\,dp \\ 
        &\propto \frac{1}{N_{{\rm MC}}} \sum\limits_{k_{{\rm MC}}}^{N_{{\rm MC}}} \frac{\left|M(p^{k_{{\rm MC}}},\alpha)\right|^{2}\gamma_{\epsilon}} { \left|M^{{\rm gen}}(p^{k_{{\rm MC}}},\alpha^\prime)\right|^{2} }. 
    \end{aligned}
\label{sig-MC-intergral-correction}
\end{equation}
    \end{linenomath}

\subsubsection{Propagator and Blatt-Weisskopf Barrier}
\label{propagator}
For a given two-body decay ($a \rightarrow bc$),  $p_{a}$, $p_{b}$ and $p_{c}$ are the momenta of particles $a$, $b$ and $c$.
The variables $s_{a}$, $s_{b}$ and $s_{c}$ refer to the squared invariant masses of particles $a$, $b$ and $c$.
The momentum $q$ is defined as the magnitude of the momentum of $b$ or $c$ in the rest system of $a$:
    \begin{linenomath}
\begin{equation}
    q=\sqrt{ \frac{(s_{a} + s_{b} - s_{c})^{2}}{4s_{a}} - s_{b}}. \label{base-q}
\end{equation}
    \end{linenomath}
The resonances $K^{*}(892)$, $f_{0}(1710)$, $\phi(1020)$ and $f_{0}(1370)$ are parameterized with a relativistic Breit-Wigner formula,
    \begin{linenomath}
\begin{equation}
    \begin{array}{lr}
        P = \frac{1}{(m_{0}^{2} - s_{a} ) - im_{0}\Gamma(m)}, &\\
        \\
        \Gamma(m) = \Gamma_{0}\left(\frac{q}{q_{0}}\right)^{2L+1}\left(\frac{m_{0}}{m}\right)\left(\frac{F_{L}(q)}{F_{L}(q_{0})}\right)^{2}, &
    \end{array}\label{RBW} 
\end{equation}
    \end{linenomath}
where $m_{0}$ and $\Gamma_{0}$ are the mass and the width of the intermediate resonance, fixed to the PDG values~\cite{PDG}, with the exception of $f_{0}(1370)$. 
The mass and width of $f_{0}(1370)$ are fixed to 1350 MeV$/c^{2}$ and 265 MeV~\cite{para-f01370}, respectively.
The value of $q_{0}$ in Eq.~(\ref{RBW}) is that of $q$ when $s_{a}=m_{0}^{2}$, $L$ denotes the angular momenta and Blatt-Weisskopf Barrier $F_{L}(q)$ is defined as:
    \begin{linenomath}
\begin{equation}
    \begin{array}{lr}
        F_{L=0}(q) = 1,       &\\
        F_{L=1}(q) = \sqrt{\frac{z_{0}^{2}+1}{z^{2}+1}},       &\\
        F_{L=2}(q) = \sqrt{\frac{z_{0}^{4}+3z_{0}^{2}+9}{z^{4}+3z^{2}+9}},       &\\
    \end{array}\label{XLQ} 
\end{equation}
    \end{linenomath}
    where $z=qR$ and $z_{0}=q_{0}R$.
$R$ is the effective radius of the intermediate resonance or $D_{s}$ meson.
The values of $R$ are fixed to $3.0\ {\rm GeV}^{-1}$ for intermediate states and $5.0\ {\rm GeV}^{-1}$  for $D_{s}$ meson, respectively.
The uncertainty of $R$ values is taken into account in evaluation 
of the systematic uncertainties.

The $K^{*}_{0}(1430)^{0}$ is parameterized with the Flatt\' e formula:
    \begin{linenomath}
\begin{equation}
    P_{K^{*}_{0}(1430)^{0}}= \frac{1}{M^{2} - s - i(g_{1}\rho_{K\pi}(s) + g_{2}\rho_{\eta^{\prime}K}(s))}, \label{Flatte}
\end{equation}
    \end{linenomath}
where $s$ is the squared $K^{-}\pi^{+}$ invariant mass,  $\rho_{K\pi}(s)$ and $\rho_{\eta^{\prime}K}(s)$ are Lorentz invariant PHSP factor, and   $g_{1,2}$ are coupling constants to the corresponding final state. 
The parameters of the $K^{*}_{0}(1430)^{0}$ are fixed to values measured by CLEO~\cite{CLEO-Flatte}. 

For the resonance $S(980)$ (representing the $f_{0}(980)$ and $a_{0}(980)$), 
we use Eq.~(\ref{S980-RBW}) to describe the propagator 
and the values of parameters are fixed to those in Eq.~(\ref{S-wave parameters}) 
obtained from the MIPWA section (Sec.~\ref{MIPWA}).

\subsubsection{Spin Factors}
\label{spin-factor}
The spin projection operators~\cite{covariant-tensors} for a two-body decay are defined as
    \begin{linenomath}
\begin{eqnarray}                                                                                                                                                                              
    \begin{aligned}
        P^{0}(a) &= 1, &\\
        P^{(1)}_{\mu\mu^{\prime}}(a) &= -g_{\mu\mu^{\prime}}+\frac{p_{a,\mu}p_{a,\mu^{\prime}}}{p_{a}^{2}}, &\\
        P^{(2)}_{\mu\nu\mu^{\prime}\nu^{\prime}}(a) &= \frac{1}{2}(P^{(1)}_{\mu\mu^{\prime}}(a)P^{(1)}_{\nu\nu^{\prime}}(a)+P^{(1)}_{\mu\nu^{\prime}}(a)P^{(1)}_{\nu\mu^{\prime}}(a)) & &\\
                                          &-\frac{1}{3}P^{(1)}_{\mu\nu}(a)P^{(1)}_{\mu^{\prime}\nu^{\prime}}(a).            &
    \end{aligned}
    \label{spin-projection-operators} 
\end{eqnarray}
    \end{linenomath}
The corresponding covariant tensors are expressed as follows
    \begin{linenomath}
\begin{equation}
    \begin{array}{lr}
        \tilde{t}^{(0)}(a) = 1, &\\
        \tilde{t}^{(1)}_{\mu}(a) = -P^{(1)}_{\mu\mu^{\prime}}(a)r^{\mu^{\prime}}_{a}, &\\
        \tilde{t}^{(2)}_{\mu\nu}(a) = P^{(2)}_{\mu\nu\mu^{\prime}\nu^{\prime}}(a)r^{\mu{\prime}}_{a}r^{\nu^{\prime}}_{a}, &\\
    \end{array}\label{covariant-tensors} 
\end{equation}
    \end{linenomath}
    where $r_{a} = p_{b} - p_{c}$ is the momentum difference between $b$ and $c$. 
The spin factor for the process $D_{s} \rightarrow aX$ (where $a$ is a resonance and $X$ is a direct daughter of the $D_{s}$ meson) with $a \rightarrow bc$ is, 
    \begin{linenomath}
\begin{equation}
    \begin{array}{lr}
        S_{n} = 1, &\\
        S_{n} = \tilde{T}^{(1)\mu}(D_{s})\tilde{t}^{(1)}_{\mu}(a), &\\
        S_{n} = \tilde{T}^{(2)\mu\nu}(D_{s})\tilde{t}^{(2)}_{\mu\nu}(a), &
    \end{array}\label{spin-factorF} 
\end{equation}
    \end{linenomath}
where $\tilde{T}^{(L)}_\mu(D_{s})$ and $\tilde{t}^{(L)}_\mu(a)$ are the covariant tensors with angular momenta $L$ for $D_{s} \rightarrow aX$ and $ a \rightarrow bc$, respectively.

\subsection{Fit Result}
\label{AA-FitResult}
We start the fit of data by considering the amplitudes containing
$\bar{K}^{*}(892)^{0}$, $\phi(1020)$ and $S(980)$ resonances, as these resonances are clearly seen in Fig.~\ref{dalitz}.
We choose $\bar{K}^{*}(892)^{0}$ as the reference amplitude and fix the magnitude  $\rho$ and  phase $\phi$ for $D_{s}^{+} \rightarrow \bar{K}^{*}(892)^{0}K^{+}$ to 1.0 and 0.0, respectively.
The magnitudes and phases of other processes are free parameters in the fit.
We then add amplitudes with resonances listed in the PDG~\cite{PDG} and non-resonant components until no additional amplitude has significance larger than 5$\sigma$.
The statistical significance for a certain intermediate process is calculated using the change of likelihood and number of degrees of freedom between with and without this process.
The six intermediate processes retained in the nominal fit are  
$D_{s}^{+} \rightarrow \bar{K}^{*}(892)^{0}K^{+}$,
$\phi(1020)\pi^{+}$,
$S(980)\pi^{+}$,
$\bar{K}^{*}_{0}(1430)^{0}K^{+}$,
$f_{0}(1370)\pi^{+}$ and
$f_{0}(1710)\pi^{+}$.
The magnitudes, phases and corresponding significances of these amplitudes are listed in Table~\ref{final-result}.
Other tested amplitudes when determining the nominal fit model, but finally not used, are listed in Table~\ref{AA-tested}.
The systematic uncertainties are discussed in Sec.~\ref{AA-sys}.
\begin{table*}[htbp]
    \caption{The results on the magnitudes, phases, FFs and significances for the six amplitudes. The first and second uncertainties are the statistical and systematic, respectively.}
    \renewcommand\arraystretch{1.2}
    \label{final-result}
    \begin{center}
        \begin{tabular}{l|cccc}
            \hline\hline
            Amplitude & Magnitude ($\rho$) & Phase ($\phi$)  & FFs (\%) & Significance ($\sigma$)\\
            \hline
            $D_{s}^{+} \rightarrow \bar{K}^{*}(892)^{0}K^{+}$              & 1.0 (fixed)             & 0.0 (fixed)                & 48.3$\pm$0.9$\pm$0.6    &  $> 20$\\
            $D_{s}^{+} \rightarrow \phi(1020)\pi^{+}$                      & 1.09$\pm$0.02$\pm$0.01 & 6.22$\pm$0.07$\pm$0.04    & 40.5$\pm$0.7$\pm$0.9      &  $> 20$ \\
            $D_{s}^{+} \rightarrow S(980)\pi^{+}$                          & 2.88$\pm$0.14$\pm$0.17 & 4.77$\pm$0.07$\pm$0.07    & 19.3$\pm$1.7$\pm$2.0      & $>20$  \\
            $D_{s}^{+} \rightarrow \bar{K}^{*}_{0}(1430)^{0}K^{+}$         & 1.26$\pm$0.14$\pm$0.16 & 2.91$\pm$0.20$\pm$0.23    & \hphantom{0}3.0$\pm$0.6$\pm$0.5   & 8.6\\
            $D_{s}^{+} \rightarrow f_{0}(1710)\pi^{+}$                     & 0.79$\pm$0.08$\pm$0.14 & 1.02$\pm$0.12$\pm$0.06    & \hphantom{0}1.9$\pm$0.4$\pm$0.6   & 9.2\\
            $D_{s}^{+} \rightarrow f_{0}(1370)\pi^{+}$                     & 0.58$\pm$0.08$\pm$0.08 & 0.59$\pm$0.17$\pm$0.46    & \hphantom{0}1.2$\pm$0.4$\pm$0.2   & 6.4\\
            \hline\hline
        \end{tabular}
    \end{center}
\end{table*}

\begin{table}[htbp]
        \caption{The significances for other tested amplitudes.}
        \renewcommand\arraystretch{1.2}
        \label{AA-tested}
        \begin{center}
            \begin{tabular}{l|c}
                \hline\hline
                Amplitude & Significance ($\sigma$) \\
                \hline
                $D_{s}^{+} \rightarrow f_{0}(1500)\pi^{+}$             & 0.8\\
                $D_{s}^{+} \rightarrow \phi(1680)\pi^{+}$              & 1.4\\
                $D_{s}^{+} \rightarrow f_{2}(1270)\pi^{+}$             & 2.5\\
                $D_{s}^{+} \rightarrow f_{2}(1525)\pi^{+}$             & 0.2\\
                $D_{s}^{+} \rightarrow \bar{K}_{1}^{*}(1410)^{0}K^{+}$ & 2.6\\
                $D_{s}^{+} \rightarrow \bar{K}_{1}^{*}(1680)^{0}K^{+}$ & 0.1\\
                $D_{s}^{+} \rightarrow \bar{K}_{2}^{*}(1430)^{0}K^{+}$ & 1.9\\
                non-resonance                                          & 3.1\\
                \hline\hline
            \end{tabular}
        \end{center}
\end{table}

\label{FF}
With the coefficients $c_{n}$ obtained from the fit, the FFs are calculated with generator-level phase-space MC as 
    \begin{linenomath}
\begin{equation}
{\rm FF}(n) = \frac{\begin{matrix}\sum \left|c_{n}A_{n}\right|^{2}\end{matrix}}{\begin{matrix}\sum \left|M(p_{j}^{k},\alpha)\right|^{2}\end{matrix}}, \label{Fit-Fraction-Definition}
\end{equation}
    \end{linenomath}
    where the summation is performed over the generated PHSP MC events. 

To properly treat correlations, we randomly vary the coefficients $c_n$ 
according to the corresponding error matrix to produce many sets of $c_n$ 
and then obtain a series of FFs for each intermediate process.  
A Gaussian function is used to fit the distribution of FF for each intermediate process 
and the width of the Gaussian function is taken as the corresponding statistical uncertainty of the FF.
The resultant FFs are listed in Table~\ref{final-result}.

\begin{figure*}[htbp]
    \centering
    \mbox{
        \begin{overpic}[width=0.8\textwidth]{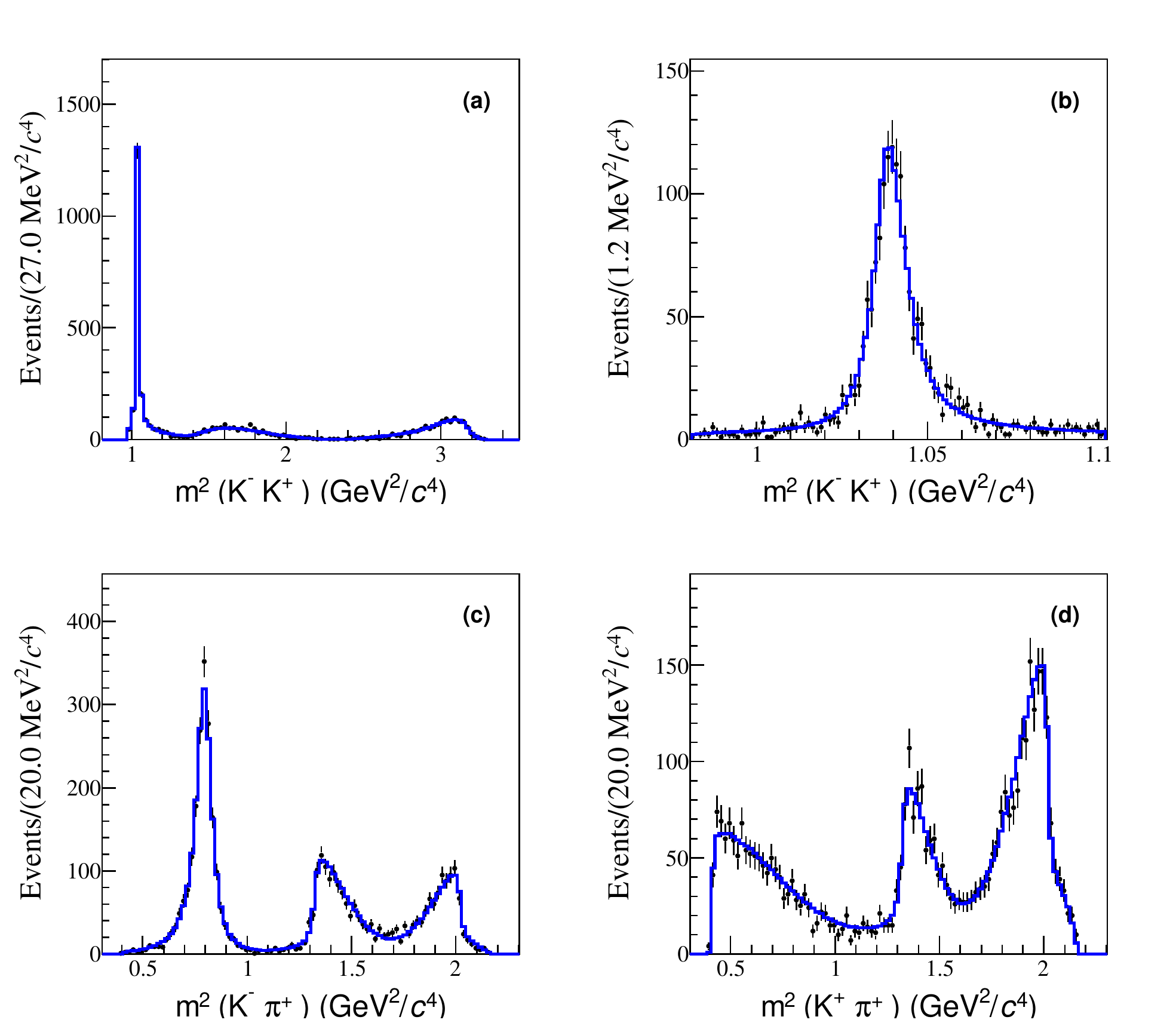}
        \end{overpic}
    }
    \caption{Dalitz plot projections (a) $m^{2}(K^{+}K^{-})$, (b) $m^{2}(K^{+}K^{-})$ near the $\phi(1020)$ peak, (c) $m^{2}(K^{-}\pi^{+})$ and (d) $m^{2}(K^{+}\pi^{+})$ from the nominal fit.
    The data are represented by points with error bars and the solid lines indicate the signal MC sample.}
    \label{dalitz-projection}
\end{figure*}

Signal MC samples modeled according to the fit result are generated to compare 
the projections of the Dalitz plots with data and to calculate the fit bias, 
which will be discussed in Sec.~\ref{AA-sys}.
The Dalitz plot projections are shown in Fig.~\ref{dalitz-projection}.
To evaluate the goodness of fit with a $\chi^{2}/{\rm NDF}$ criterion, we calculate $\chi^{2} = \sum{\left(\frac{ N_{{\rm data}} - N_{{\rm exp}}}{\sigma_{{\rm data}}}\right)}^{2}$ of the fit using an adaptive binning of the Dalitz plot of $m^{2}(K^{+}K^{-})$ versus $m^{2}(K^{-}\pi^{+})$, in which each bin has at least 10 events.
Here $N_{{\rm data}}$, $\sigma_{{\rm data}}$ and $N_{{\rm exp}}$ refer to the number of events from data, the error of $N_{{\rm data}}$ and the
expected number obtained from signal MC in each bin, respectively.  
We find a $\chi^{2}/{\rm NDF} = 290.0/280$.

\subsection{Systematic Uncertainty}
\label{AA-sys}
The following categories of systematic uncertainties are studied for the amplitude analysis: 

\begin{enumerate} 
    \item [\uppercase\expandafter{\romannumeral1}] Resonance parameters. 
        The masses and widths of resonances are shifted by their corresponding uncertainties. 
        For the $S(980)$, $m_{0}$ and $\Gamma_{0}$ are shifted according to the errors from Eq.~(\ref{S-wave-sys}).
        The mass and width of $f_{0}(1370)$ are shifted according to the uncertainties from Ref.~\cite{para-f01370}.
        The parameters of $\bar{K}^{*}_{0}(1430)^{0}$ are shifted according to the errors from Ref.~\cite{CLEO-Flatte}.
        For other states, uncertainties are taken from the PDG~\cite{PDG}.
    \item [\uppercase\expandafter{\romannumeral2}] The effective radius in the Blatt-Weisskopf barrier factor is varied within the range $\left[1.0, 5.0\right] \ {\rm GeV}^{-1}$ for intermediate resonances and  $\left[3.0, 7.0\right] \ {\rm GeV}^{-1}$ for $D_{s}$ mesons. 
    \item [\uppercase\expandafter{\romannumeral3}] Fit bias. 
        Pull distribution checks using 300 signal MC samples are performed to obtain the fit bias. 
        Here the pull value for a certain parameter $x$ is defined as $(x_{{\rm true}} - x_{{\rm MC}})/\sigma_{x_{{\rm MC}}}$,
        where $x_{{\rm MC}}$ and $\sigma_{x_{{\rm MC}}}$ are the value and the statistical error of $x$ obtained from the fit to a certain signal MC sample and $x_{{\rm true}}$ refers to the true value of $x$ used in the MC generation.
        The signal MC samples each have the same size as the data. 
        Fits to the pull distributions with Gaussian functions show no obvious biases and under- or over-estimations on statistical uncertainties.
        We add quadrature sum of the mean value and the error of mean to get the corresponding systematic uncertainty in units of the corresponding statistical uncertainty.
        \\ \\
    \item [\uppercase\expandafter{\romannumeral4}] Detector effects. 
        These effects are related to the efficiency difference between MC simulation and data caused by PID and tracking, reflected in the $\gamma_{\epsilon}$ in Eq.~(\ref{experimental-effect}).
        The uncertainties associated with $\gamma_{\epsilon}$ are obtained by performing alternative amplitude analyses varying PID and tracking efficiencies according to their uncertainties.
    \item [\uppercase\expandafter{\romannumeral5}] Model assumptions. 
        We replace the Flatt\' e expression in Eq.~(\ref{Flatte}) with  the LASS model~\cite{LASS}. 
        For the $S(980)$, Eq.~(\ref{S980-RBW}) is replaced with the Flatt\' e parameterization~\cite{Flatte} to describe the lineshape of the $S(980)$ and the parameters in the Flatt\' e parameterization are obtained from the fit to $\left|S\right|^{2}$ in Fig.~\ref{SP}(a).  
        The quadrature sum of the shifts in the results are taken as the corresponding systematic uncertainties.
    
    \item [\uppercase\expandafter{\romannumeral6}] Background estimation.
        The background is ignored in the nominal fit.
        We subtract the contribution of the background by assigning a negative weight to the background events in the likelihood calculation~\cite{bkg}.
        Individual changes of the results with respect to the nominal one are taken as the corresponding systematic
        uncertainties.
    \item [\uppercase\expandafter{\romannumeral7}] Contributions with statistical significances less than 5$\sigma$.
        The intermediate processes with statistical significances less than 5$\sigma$ are added in the nominal fit one by one.
        The quadrature sum of each parameter variations is taken as the corresponding systematic uncertainty.
\end{enumerate}

Systematic uncertainties on the magnitudes, phases and FFs are summarized 
in Table~\ref{systematic-uncertainties} and the total uncertainties 
are obtained as the sum of all the contributions in quadrature.

\begin{table*}[htbp]  
        \centering  
        \caption{Systematic uncertainties on the $\phi$, $\rho$ and FFs for different amplitudes in units of the corresponding statistical uncertainties.
        Here \uppercase\expandafter{\romannumeral1},  \uppercase\expandafter{\romannumeral2}, \uppercase\expandafter{\romannumeral3}, \uppercase\expandafter{\romannumeral4}, \uppercase\expandafter{\romannumeral5}, \uppercase\expandafter{\romannumeral6} and \uppercase\expandafter{\romannumeral7} 
        denote the propagator parameterizations of the resonances, the effective radius of Blatt-Weisskopf barrier factor, fit bias, detector effects, model assumptions, background estimation and contributions with statistical significances less than 5$\sigma$, respectively.  
        The quadrature sums of these terms are taken as the total systematic uncertainties.
    }  
        \label{systematic-uncertainties}  
        \renewcommand\arraystretch{1.2}
        \begin{tabular}{l|ccccccccc} 
            \hline\hline
            \multirow{2}{*}{Amplitude }&\multicolumn{9}{c}{Source}\cr 
            & & \uppercase\expandafter{\romannumeral1} &\uppercase\expandafter{\romannumeral2} &\uppercase\expandafter{\romannumeral3} &\uppercase\expandafter{\romannumeral4} &\uppercase\expandafter{\romannumeral5}&\uppercase\expandafter{\romannumeral6}&\uppercase\expandafter{\romannumeral7}& Total   \\
            \hline
            $D_{s}^{+} \rightarrow \bar{K}^{*}(892)^{0}K^{+}$                           &FF             &0.32      &0.29       &0.14   &0.41  &0.14  &0.10  &0.14  &0.65   \\
            \hline                                                                                                                                                        
            \multirow{3}{*}{$D_{s}^{+} \rightarrow \phi(1020)\pi^{+}$}                  & $\phi$        &0.49      &0.10       &0.06   &0.07  &0.08  &0.02  &0.04  &0.52 \\
                                                                                        & $\rho$        &0.49      &0.14       &0.08   &0.41  &0.19  &0.01  &0.17  &0.71 \\
                                                                                        & FF            &0.44      &1.13       &0.04   &0.40  &0.08  &0.01  &0.15  &1.29 \\
            \hline                                                                                                                                                       
            \multirow{3}{*}{$D_{s}^{+} \rightarrow S(980)\pi^{+}$}                      & $\phi$        &0.98      &0.25       &0.04   &0.11  &0.06  &0.03  &0.15  &1.03    \\
                                                                                        & $\rho$        &1.11      &0.17       &0.09   &0.11  &0.20  &0.17  &0.23  &1.18 \\
                                                                                        & FF            &1.16      &0.15       &0.04   &0.09  &0.05  &0.04  &0.25  &1.20 \\
            \hline                                                                                                                                                       
            \multirow{3}{*}{$D_{s}^{+} \rightarrow \bar{K}^{*}_{0}(1430)^{0}K^{+}$}     & $\phi$        &1.02      &0.48       &0.05   &0.21  &0.09  &0.06  &0.16  &1.16     \\
                                                                                        & $\rho$        &1.00      &0.36       &0.15   &0.20  &0.19  &0.02  &0.21  &1.12 \\
                                                                                        & FF            &0.76      &0.35       &0.11   &0.22  &0.19  &0.03  &0.20  &0.92 \\
            \hline                                                                                                                                                       
            \multirow{3}{*}{$D_{s}^{+} \rightarrow f_{0}(1710)\pi^{+}$}                 & $\phi$        &0.31      &0.25       &0.04   &0.14  &0.17  &0.01  &0.17  &0.49 \\
                                                                                        & $\rho$        &1.17      &1.23       &0.09   &0.11  &0.11  &0.11  &0.01  &1.71 \\
                                                                                        & FF            &0.71      &1.21       &0.04   &0.16  &0.10  &0.10  &0.01  &1.42 \\
            \hline                                                                                                                                                       
            \multirow{3}{*}{$D_{s}^{+} \rightarrow f_{0}(1370)\pi^{+}$}                 & $\phi$        &2.66      &0.27       &0.12   &0.09  &0.28  &0.21  &0.20  &2.71  \\
                                                                                        & $\rho$        &1.01      &0.32       &0.21   &0.09  &0.05  &0.03  &0.21  &1.10 \\
                                                                                        & FF            &0.42      &0.30       &0.15   &0.06  &0.15  &0.09  &0.19  &0.60 \\
            \hline\hline
        \end{tabular}  
    \end{table*}

\section{Branching Fraction Measurement}
\label{BF}
\subsection{Efficiency and Data Yields}
\begin{figure*}[!htbp]
 \centering
 \includegraphics[width=0.35\textwidth]{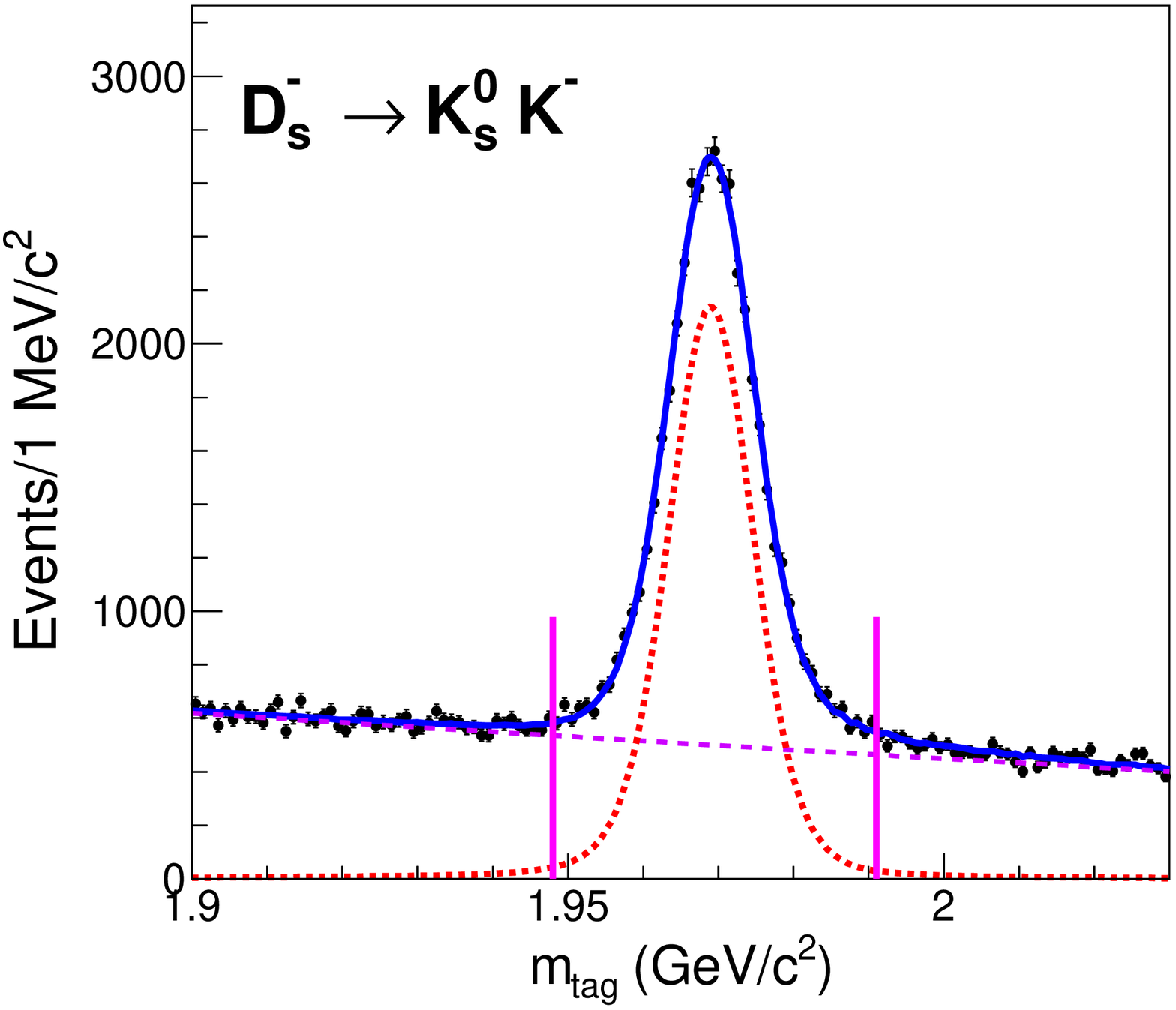}
 \includegraphics[width=0.35\textwidth]{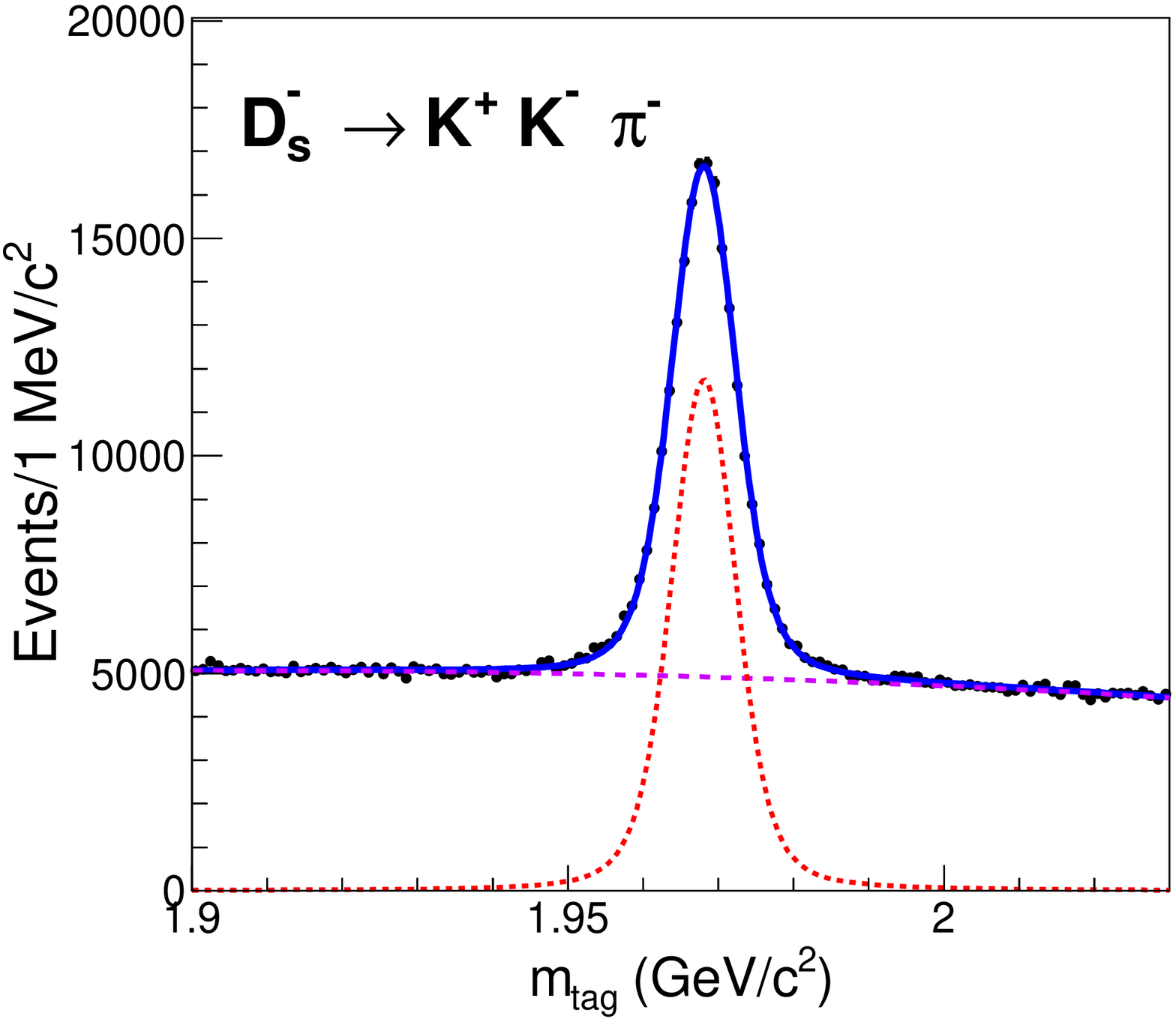}
 \includegraphics[width=0.35\textwidth]{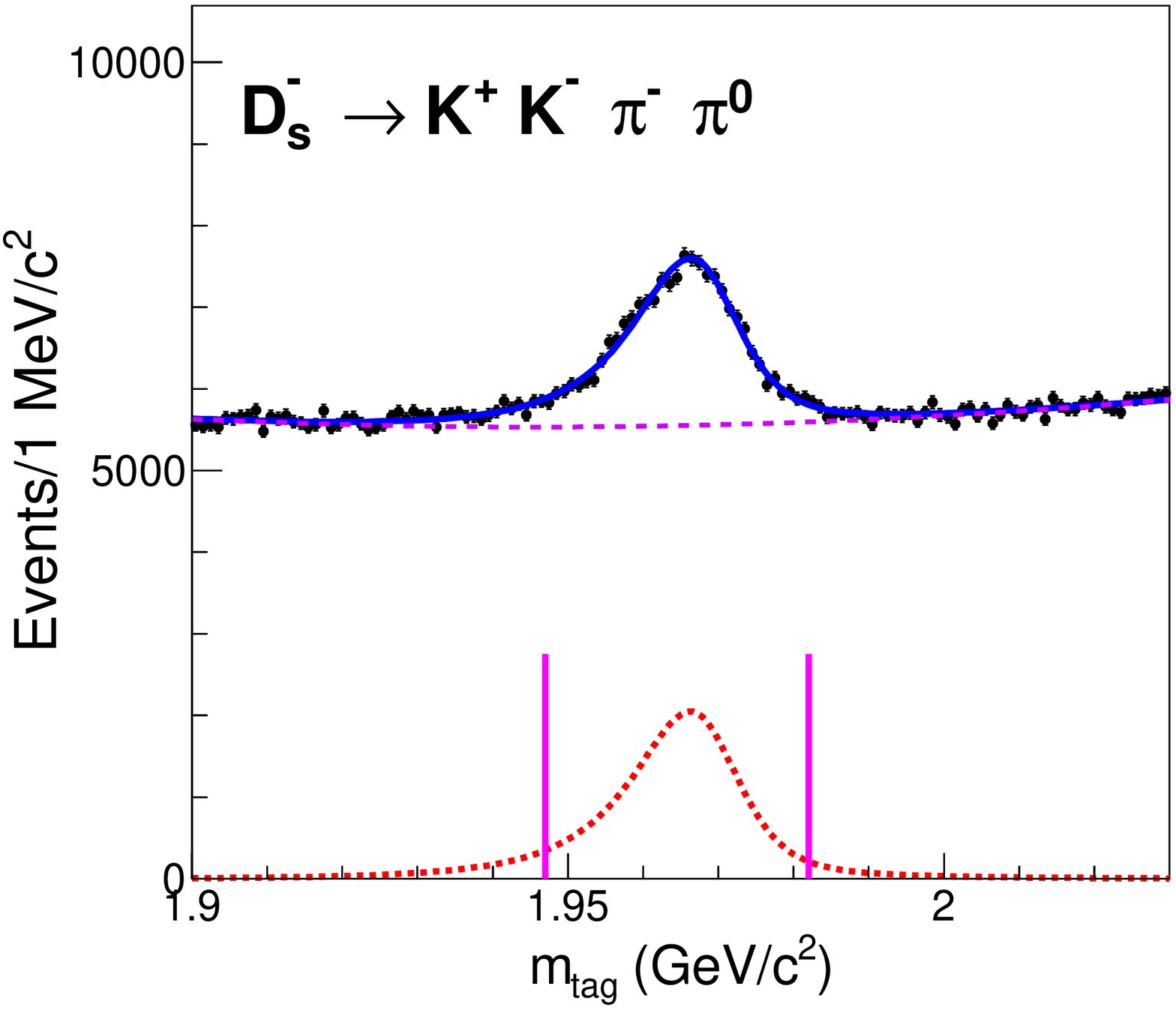}
 \includegraphics[width=0.35\textwidth]{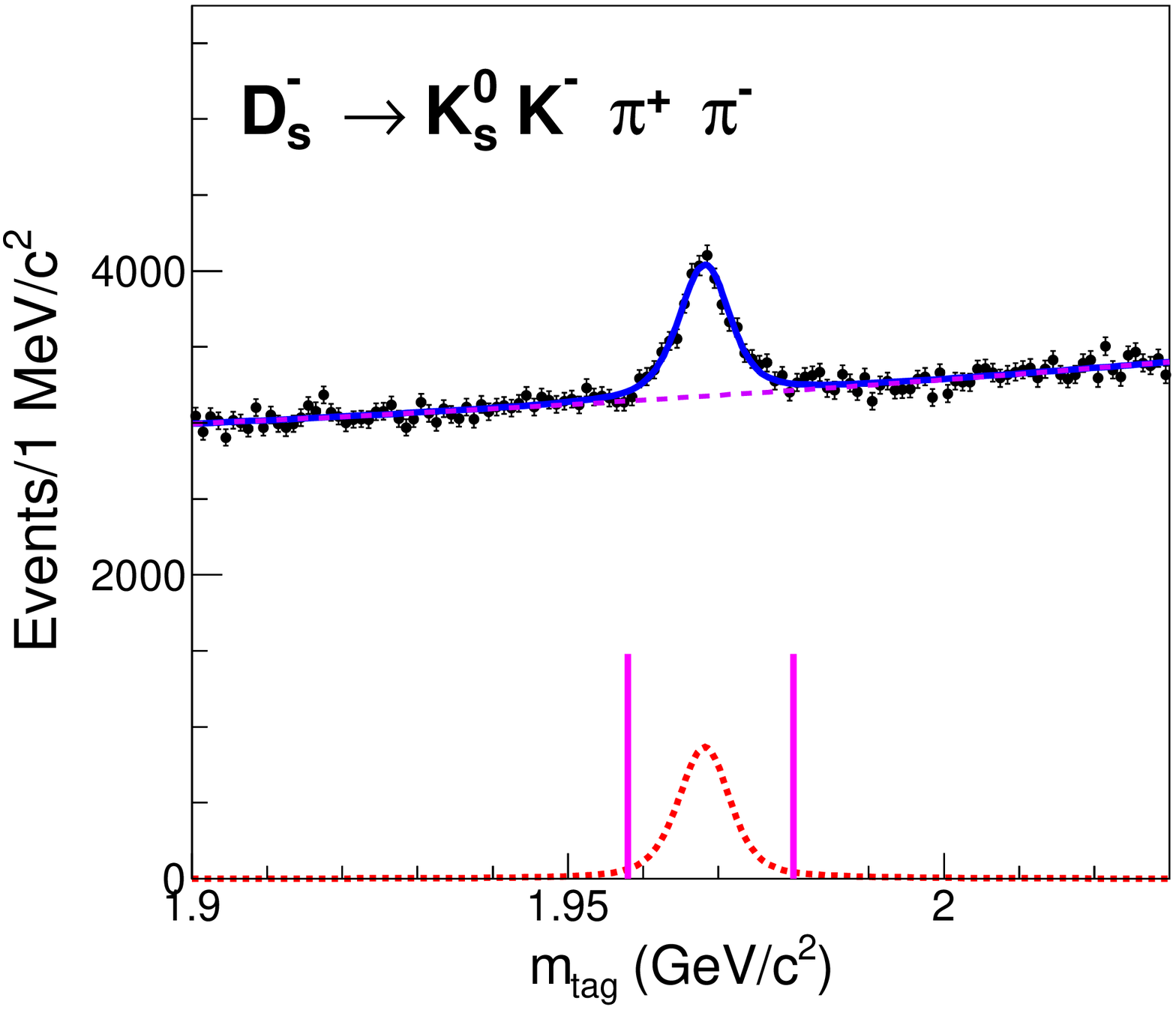}
 \includegraphics[width=0.35\textwidth]{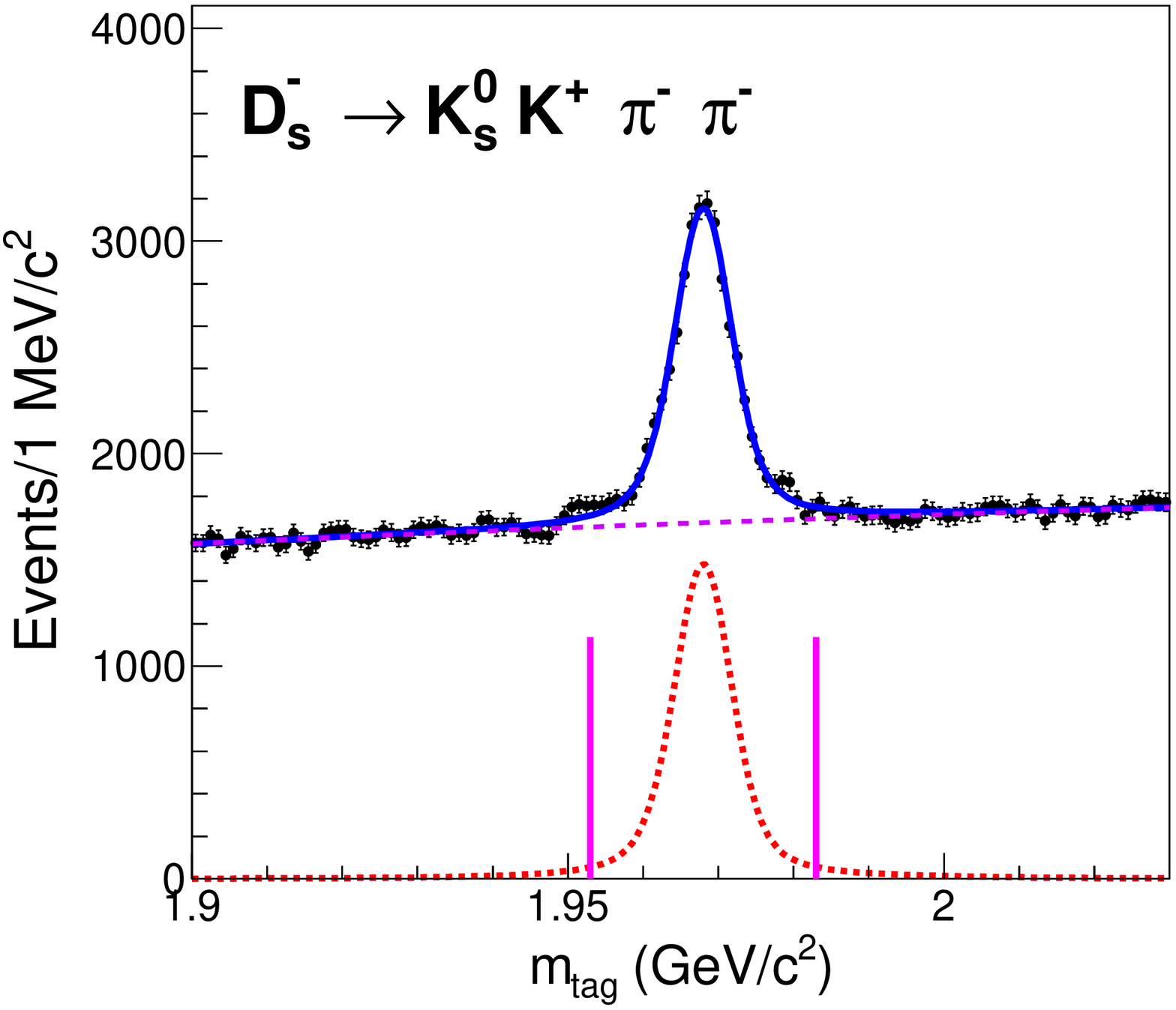}
 \includegraphics[width=0.35\textwidth]{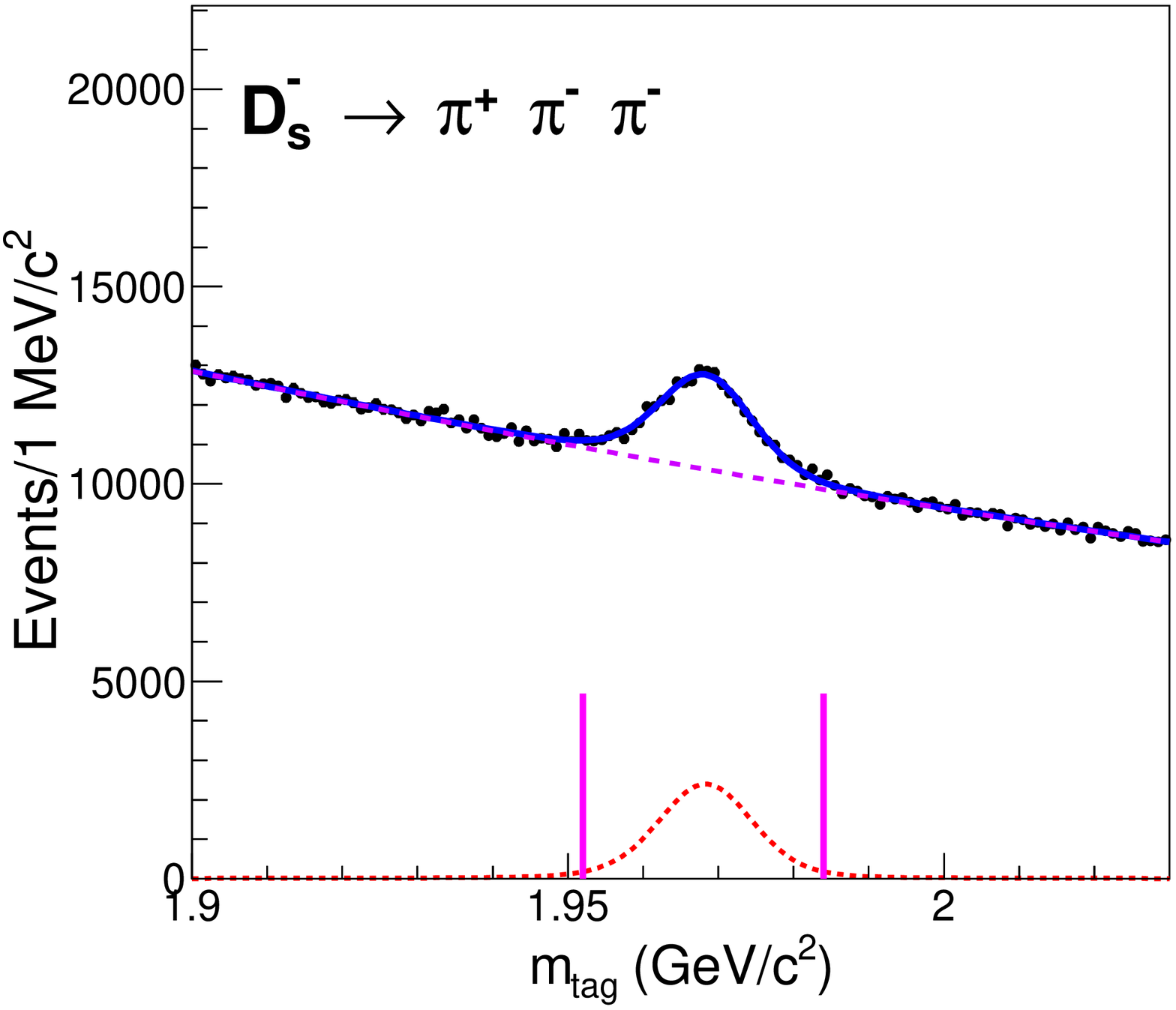}
 \includegraphics[width=0.35\textwidth]{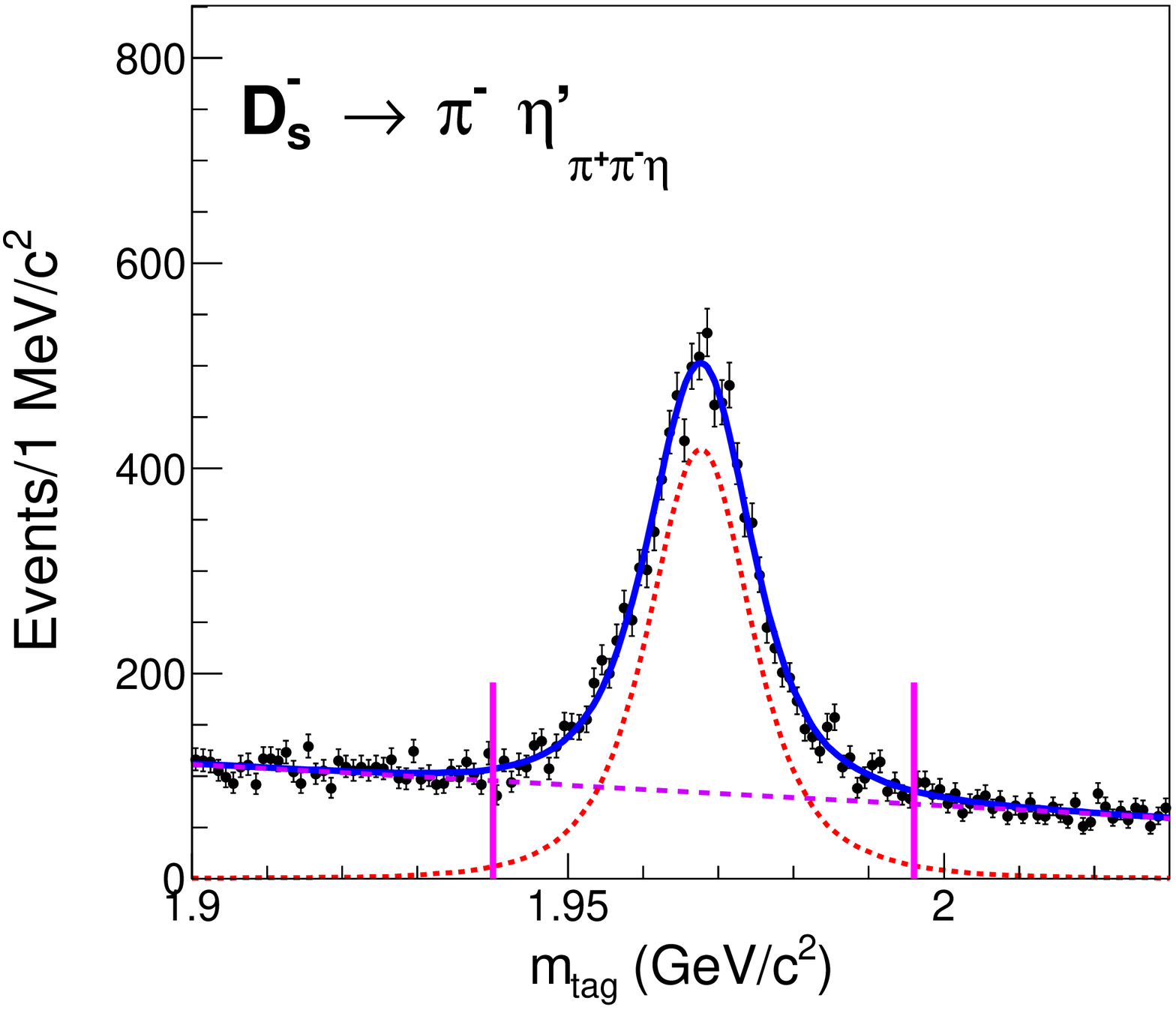}
 \includegraphics[width=0.35\textwidth]{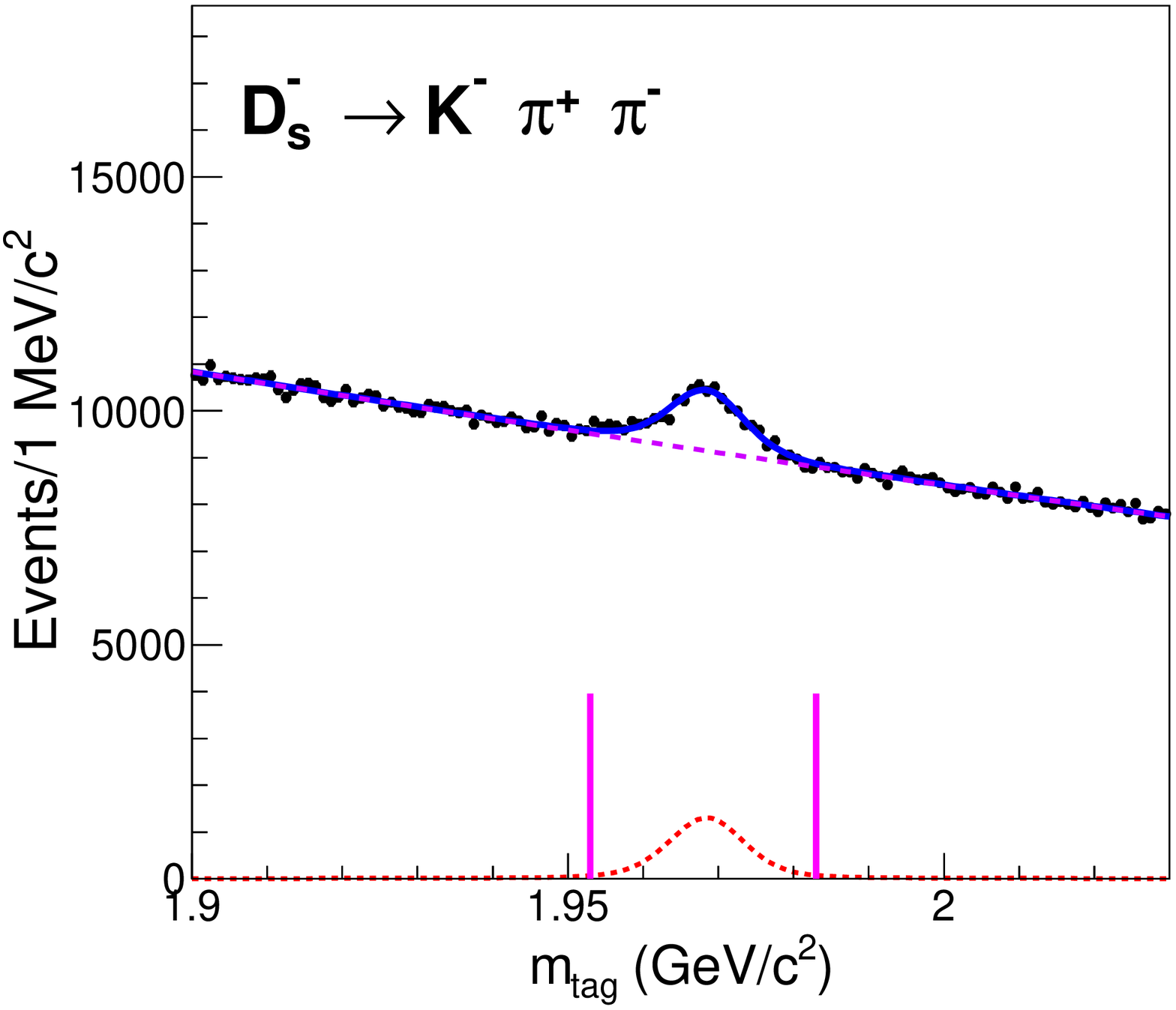}
 \caption{Fits to the $m_{{\rm tag}}$ distributions of data. 
 The points with error bars indicate data and the solid lines indicate the fit. 
 Red short-dashed lines are signal, violet long-dashed lines are background. 
 The region within the purple lines denotes the signal region. 
 }
\label{SingleTagFit}
\end{figure*}

After the selection described in Sec.~\ref{chap:event_selection}, the tag technique is also used to perform the BF measurement.
We use the same eight tag modes as in Sec.~\ref{AA}.
For each tag mode, if there are multiple tag $D_{s}$ candidates in an event, the candidate with $M_{{\rm rec}}$ closest to the nominal mass of $D_{s}^{*}$~\cite{PDG} is retained.
The ST yields are obtained by the fits to the $D_{s}$ invariant mass distributions, as shown in Fig.~\ref{SingleTagFit}, along with the mass windows 
listed in Table~\ref{ST-mass-window}.  
The signal shape is modeled as the MC-simulated shape convolved with a Gaussian function, while background is parameterized as a second-order Chebychev polynomial.
Fits to $m_{{\rm tag}}$ for inclusive MC are performed to estimate the corresponding ST efficiencies.  
The ST yields ($Y_{{\rm ST}}$) and ST efficiencies ($\epsilon_{{\rm ST}}$) are listed in Table~\ref{ST-eff}.

\begin{table*}[htbp]
    \caption{ The ST yields ($Y_{{\rm ST}}$) and ST efficiencies ($\epsilon_{{\rm ST}}$). 
The BFs of the sub-particle ($K_{S}^{0}$, $\pi^{0}$, $\eta$ and $\eta^{\prime}$) decays are not included in the efficiencies.}
    \label{ST-eff}
    \begin{center}
        \begin{tabular}{l|crc}
            \hline\hline
            Tag mode & Mass window (GeV/$c^{2}$)  & $Y_{{\rm ST}}\hphantom{0000}$  & $\epsilon_{{\rm ST}}(\%)$\\
            \hline
            $D_{s}^{-} \rightarrow K_{S}^{0}K^{-}$                          & [1.948, 1.991]    & \hphantom{0}31987$\pm$\hphantom{0}314                     & 47.66$\pm$0.07\\
            $D_{s}^{-} \rightarrow K^{+}K^{-}\pi^{-}$                       & -                 &            141189$\pm$\hphantom{0}643                     & 40.90$\pm$0.03\\
            $D_{s}^{-} \rightarrow K^{+}K^{-}\pi^{-}\pi^{0}_{\gamma\gamma}$ & [1.947, 1.982]    & \hphantom{0}37899$\pm$1739                     & 10.36$\pm$0.03\\
            $D_{s}^{-} \rightarrow K_{S}^{0}K^{-}\pi^{+}\pi^{-}$            & [1.958, 1.980]    & \hphantom{00}7999$\pm$\hphantom{0}236                     & 18.67$\pm$0.12\\
            $D_{s}^{-} \rightarrow K_{S}^{0}K^{+}\pi^{-}\pi^{-}$            & [1.953, 1.983]    & \hphantom{0}15723$\pm$\hphantom{0}290                     & 21.51$\pm$0.06\\
            $D_{s}^{-} \rightarrow \pi^{-}\pi^{-}\pi^{+}$                   & [1.952, 1.984]    & \hphantom{0}38157$\pm$\hphantom{0}873                     & 50.05$\pm$0.15\\
            $D_{s}^{-} \rightarrow \pi^{-}\eta_{\pi^{+}\pi^{-}\eta_{\gamma\gamma}}^{\prime}$    & [1.940, 1.996]    & \hphantom{00}8009$\pm$\hphantom{0}142 & 19.43$\pm$0.06\\
            $D_{s}^{-} \rightarrow K^{-}\pi^{+}\pi^{-}$                     & [1.953, 1.983]    & \hphantom{0}17112$\pm$\hphantom{0}561                     & 45.66$\pm$0.22\\
            \hline\hline
        \end{tabular}
    \end{center}
\end{table*}

After the best candidates of ST $D_{s}^{-}$ mesons are identified, we search for the $D_{s}^{+} \rightarrow K^{+}K^{-}\pi^{+}$.
Only the best $D_{s}^{+}$ candidate with the average mass of tag $D_{s}^-$ and signal $D_{s}^{+}$ closest to the nominal mass of $D_{s}$ is retained for each tag mode in an event.
For all tag modes, we found a mean of 0.16\% (0.17\%) of DT events in data (inclusive MC) contains multiple $D_s^+$ candidates.The effect due to the multiple candidate selection is, therefore, negligible.
The DT efficiencies, listed in Table~\ref{DT-eff}, are obtained based on the signal MC samples.

\begin{table}[htbp]
    \caption{ The DT efficiencies ($\epsilon_{{\rm DT}}$).The BFs of the sub-particle ($K_{S}^{0}$, $\pi^{0}$, $\eta$ and $\eta^{\prime}$) decays are not included in the efficiencies.}
    \label{DT-eff}
    \begin{center}
        \begin{tabular}{l|ccc}
            \hline\hline
            Tag mode   & $\epsilon_{{\rm DT}}(\%)$\\
            \hline
            $D_{s}^{-} \rightarrow K_{S}^{0}K^{-}$                                                   & 18.59$\pm$0.14\\
            $D_{s}^{-} \rightarrow K^{+}K^{-}\pi^{-}$                                                & 17.41$\pm$0.06\\
            $D_{s}^{-} \rightarrow K^{+}K^{-}\pi^{-}\pi^{0}$                                         & \hphantom{0}4.33$\pm$0.03\\
            $D_{s}^{-} \rightarrow K_{S}^{0}K^{-}\pi^{+}\pi^{-}$                                     & \hphantom{0}8.03$\pm$0.11\\
            $D_{s}^{-} \rightarrow K_{S}^{0}K^{+}\pi^{-}\pi^{-}$                                     & \hphantom{0}8.25$\pm$0.09\\
            $D_{s}^{-} \rightarrow \pi^{-}\pi^{-}\pi^{+}$                                            & 20.84$\pm$0.13\\
            $D_{s}^{-} \rightarrow \pi^{-}\eta_{\pi^{+}\pi^{-}\eta_{\gamma\gamma}}^{\prime}$               &  \hphantom{0}8.30$\pm$0.11\\
            $D_{s}^{-} \rightarrow K^{-}\pi^{+}\pi^{-}$                                              & 19.07$\pm$0.13\\
            \hline\hline
        \end{tabular}
    \end{center}
\end{table}

\begin{figure}[!htbp]
    \centering
    \includegraphics[width=0.4\textwidth]{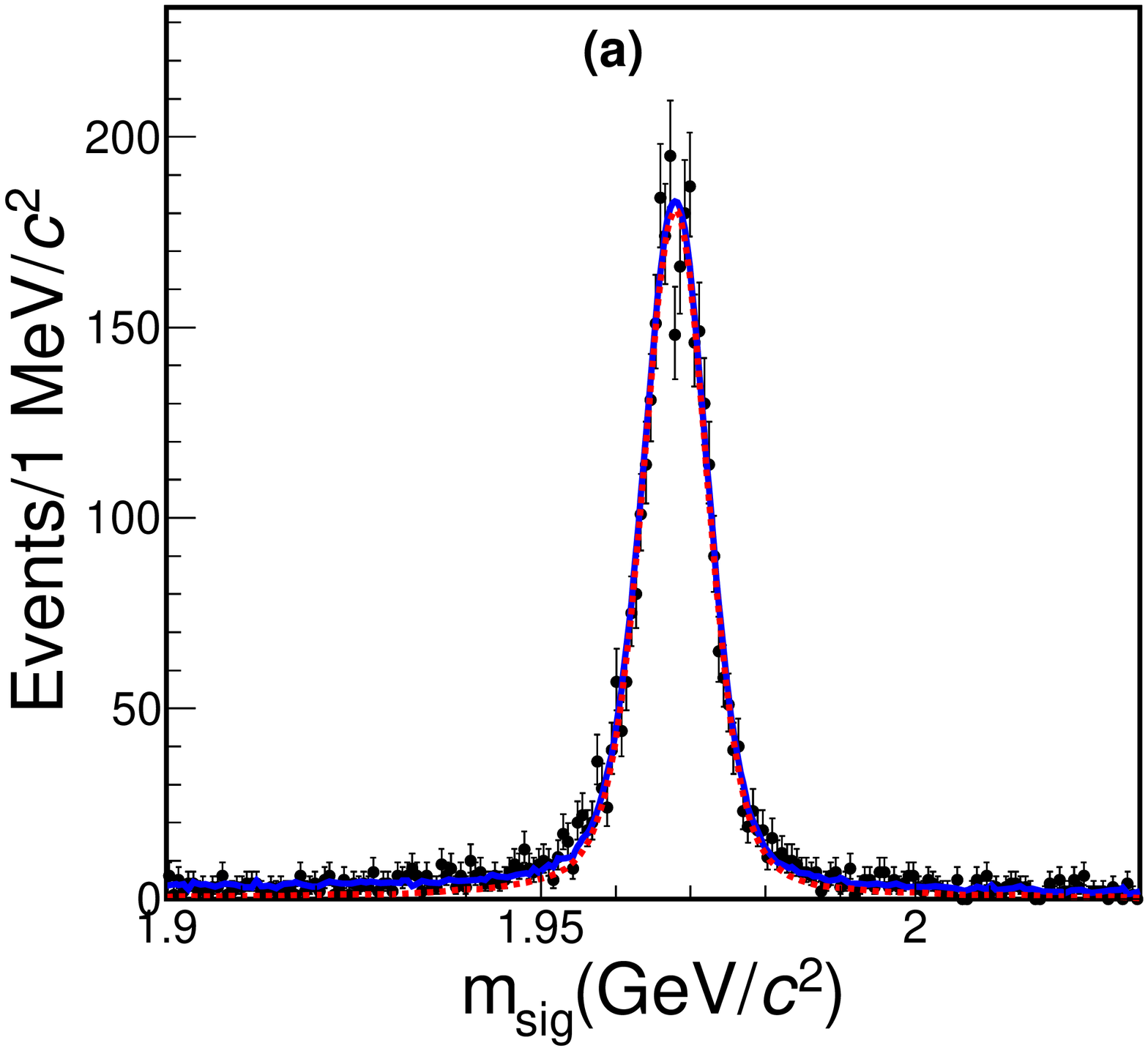}
    \includegraphics[width=0.4\textwidth]{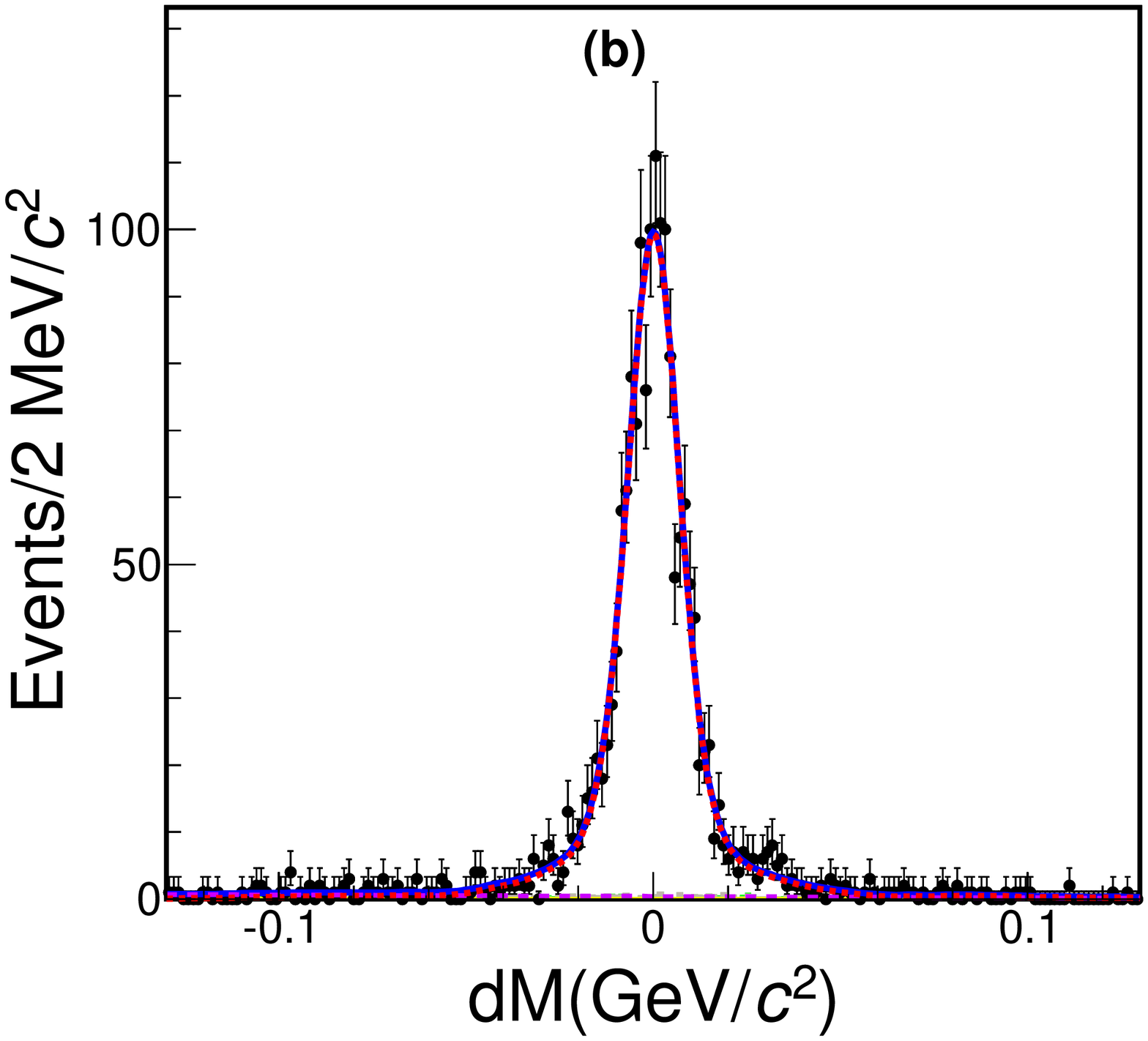}
    \caption{Fit of $m_{{\rm sig}}$ for (a) Cat.~A and $dM$ for (b) Cat.~B.
        The signal shapes are the corresponding simulated shapes convolved with a Gaussian function and 
    the background shapes are described with second-order Chebychev polynomials.}
    \label{DT-fit}
\end{figure}

As $D_{s}^{-} \rightarrow K^{+}K^{-}\pi^{-}$ is not only the signal mode but also one of the tag modes, we divide the events into two categories:

\begin{itemize}
    \item[-] Cat.~A: Tag $D_{s}^{-}$ decays to one of the tag modes except $D_{s}^{-} \rightarrow K^{+}K^{-}\pi^{-}$. The inclusive MC sample with the signal removed shows no peaking background around the fit range of $1.90 < m_{{\rm sig}} < 2.03 \ {\rm GeV}/c^{2}$.
        Thus, the DT yield is determined by the fit to $m_{{\rm sig}}$, shown in Fig.~\ref{DT-fit}(a). The background is described with a second-order Chebychev polynomial. 
        The DT yield is $3497\pm64$. 
    \item[-] Cat.~B: Tag $D_{s}^{-}$ decays to $K^{+}K^{-}\pi^{-}$. As both of the two $D_{s}$ mesons decay to the signal modes, we fit $dM$ (the mass of the signal  $D_{s}^{+}$ minus that of the tag $D_{s}^{-}$), which is shown in Fig.~\ref{DT-fit}(b). 
        Here, the background is described by a second-order Chebychev polynomial. 
        The DT yield is $1651\pm42$. 
\end{itemize}

\subsection{Tagging Technique and Branching Fraction}
For the DT samples with a certain tag mode $\alpha$, we have
    \begin{linenomath}
\begin{equation}
    Y_{{\rm ST}}^{\alpha} = 2N_{D_{s}^{+}D_{s}^{-}}\mathcal{B}_{{\rm tag}}^{\alpha}\epsilon_{{\rm tag}}^{\alpha}, \label{eq-ST}
\end{equation}
    \end{linenomath}
and
    \begin{linenomath}
\begin{equation}
    \begin{array}{lr}
        N_{{\rm sig}}^{{\rm obsA},\alpha}=2N_{D_{s}^{+}D_{s}^{-}}\mathcal{B}_{{\rm tag}}^{\alpha}\mathcal{B}_{{\rm sig}}\epsilon_{{\rm tag,sig}}^{\alpha}  , &\text{for Cat.~A} \\
        N_{{\rm sig}}^{{\rm obsB}, \alpha}=N_{D_{s}^{+}D_{s}^{-}}\mathcal{B}_{{\rm tag}}^{\alpha}\mathcal{B}_{{\rm sig}}\epsilon_{{\rm tag,sig}}^{\alpha}  ,  &\text{  for Cat.~B}  
    \end{array}
    \label{eq-DT}
\end{equation}
    \end{linenomath}
where $N_{D_{s}^{+}D_{s}^{-}}$ is the total number of $D_{s}^{*\pm}D_{s}^{\mp}$ produced from $e^{+}e^{-}$ collision;
the yields $N_{{\rm sig}}^{{\rm obsA}, \alpha}$ and $N_{{\rm sig}}^{{\rm obsB}, \alpha}$ refer to the yields with tag mode $\alpha$ for Cat.~A and Cat.~B, respectively;
$\mathcal{B}_{{\rm tag}}$ and $\mathcal{B}_{{\rm sig}}$ are the BFs of a specific tag mode and the signal mode, respectively; 
$\epsilon_{{\rm tag}}$ is the efficiency to reconstruct the tag mode; $\epsilon_{{\rm tag,sig}}$ is the efficiency to reconstruct both the tag and signal decay modes.

Using the above equations, one can obtain:
\begin{linenomath}
    \begin{equation}
    \mathcal{B}_{{\rm sig}} = \frac{N_{{\rm sig}}^{{\rm obsA}}+2N_{{\rm sig}}^{{\rm obsB}}}{\begin{matrix}\sum\limits_{\alpha} Y_{{\rm ST}}^{\alpha}\epsilon_{{\rm tag,sig}}^{\alpha}/\epsilon_{{\rm tag}}^{\alpha}\end{matrix}}, \label{BR-formula}
    \end{equation}
\end{linenomath}
where the yields $N_{{\rm sig}}^{{\rm obsA}}$, $N_{{\rm sig}}^{{\rm obsB}}$ and $Y_{{\rm ST}}^{\alpha}$ are obtained from data, while $\epsilon_{{\rm tag}}$ and $\epsilon_{{\rm tag,sig}}$ can be obtained from the updated inclusive MC samples.
The process $D_{s}^{+} \rightarrow K^{+}K^{-}\pi^{+}$ in the updated inclusive MC is generated with the Dalitz model obtained in Sec.~\ref{AA}.

\subsection{Systematic Uncertainty}
\label{BF-sys}
Most systematic uncertainties related to the efficiency for reconstructing the tag side cancel for BF measurement
due to the DT technique.
The following sources are taken in account to calculate systematic uncertainty.

\begin{itemize}
    \item Uncertainty in the number of ST $D_{s}^{-}$ candidates. 
        We perform alternative fits with different background shapes and signal 
        shapes to obtain these uncertainties.  
        We change the background shape from a second-order Chebychev polynomial to a third-order Chebychev polynomial and the relative change of BF is 0.18\%.
        The systematic uncertainty in signal shape is determined to be 0.16\% by performing an alternative fit without convolution with the Gaussian smearing function.
        The quadrature sum of these terms, that is the uncertainty in the number of ST $D_{s}^{-}$ candidates, is 0.23\%. 

    \item DT signal shape. The systematic uncertainty due to the signal shape is studied with the fit without the Gaussian function convolved, the DT yield shift is taken as the related uncertainty. 

    \item DT background shape. For background shape in the fit, a third-order Chebychev polynomial is used to replace the nominal one. 
        The quadrature sum of the BF shifts is taken as the related uncertainty.

    \item Fit bias. 
        The updated inclusive MC samples are used as fake data to estimate the possible fit bias. 
        The BF for each sample is determined and 
        the relative difference between the average of BFs and the MC truth value is $0.1\%$, which is negligible.

    \item $K^{\pm}$ and $\pi^{\pm}$ tracking/PID efficiency. 
        The ratios between data and MC efficiencies are weighted by the corresponding momentum spectra of signal MC events.
        We obtain the systematic uncertainties related to tracking efficiency to be 0.5\% for each kaon track and 0.2\% for each pion track based on the study of the tracking efficiency.
        The systematic uncertainties related to PID efficiencies are estimated to be 0.5\% for each $K^{\pm}$ and  0.4\% for each $\pi^{\pm}$.
        Tracking efficiency systematics are added linearly for the three tracks, as are the PID efficiency systematics.  

\item MC statistics. The uncertainty due to the MC statistics is obtained as $\sqrt{ \begin{matrix} \sum\limits_{\alpha}
{\left(f_{\alpha}\frac{\delta_{\epsilon_{\alpha}}}{\epsilon_{\alpha}}\right)}^{2}\end{matrix}}$, where $f_{\alpha}$ is the DT yield fraction, $\epsilon_{\alpha}$ is the DT signal efficiency of the tag mode $\alpha$ and $\delta_{\epsilon_{\alpha}}$ is the error on $\epsilon_{\alpha}$ due to the limited MC statistics.
    
    \item Dalitz model. The uncertainty from the Dalitz model is estimated as the change of efficiency when the Dalitz model parameters ($c_{n}$) are varied according to the error matrix.
\end{itemize}

All of the systematic uncertainties mentioned above are summarized in Table~\ref{BF-Sys}.
We take the quadrature sum of the systematic uncertainties above as the total systematic uncertainty in the BF of $D_{s}^{+} \rightarrow K^{+}K^{-}\pi^{+}$.
\begin{table}[htbp]
    \caption{The relative systematic uncertainties on the BF.
    The quadrature sum of all contributions is taken as the total uncertainty.
    }
    \label{BF-Sys}
    \begin{center}
        \begin{tabular}{l|ccc}
            \hline\hline
            Source   & Sys. Uncertainty (\%)\\
            \hline
            Number of $D_{s}^{-}$               & 0.2 \\
            Signal shape                        & 0.5 \\
            Background shape       & 0.9 \\
            Fit bias                & 0.1 \\
            $K^{\pm}$ and $\pi^{\pm}$ Tracking efficiency       & 1.2 \\
            $K^{\pm}$ and $\pi^{\pm}$ PID efficiency            & 1.4 \\
            MC statistics                       & 0.6 \\
            Dalitz model                               & 0.5 \\
            \hline
            Total                               & 2.3 \\
            \hline\hline
        \end{tabular}
    \end{center}
\end{table}


\section{Conclusion}
\label{CONCLUSION}
This paper presents the amplitude analysis of the decay $D_{s}^{+} \rightarrow K^{+}K^{-}\pi^{+}$.
The results on FFs for $D_{s}^{+} \rightarrow f_{0}(1370)\pi^{+}$, $D_{s}^{+} \rightarrow f_{0}(1710)\pi^{+}$ and $D_{s}^{+} \rightarrow f_{0}(980)\pi^{+}/a_{0}(980)\pi^{+}$ are consistent with those of BaBar and E687.
In addition, our results on FFs also agree with those of CLEO, except for $D_{s}^{+} \rightarrow f_{0}(980)\pi^{+}/a_{0}(980)\pi^{+}$ and  $D_{s}^{+} \rightarrow f_{0}(1370)\pi^{+}$ where 2.4$\sigma$ and 3.4$\sigma$ differences, respectively, with CLEO are observed.

    In this analysis, as $a_{0}(980)$ and $f_{0}(980)$ overlap and parameters of $a_{0}(980)$ and $f_{0}(980)$ are not well measured, 
    we have extracted the S-wave lineshape in the low end of $K^{+}K^{-}$ mass spectrum with a model-independent method.

    We have also measured the BF $\mathcal{B}(D_{s}^{+} \rightarrow K^{+}K^{-}\pi^{+})=(5.47\pm0.08_{{\rm stat}}\pm0.13_{{\rm sys}})\%$
    which is currently the most precise measurement. Comparisons
with other results are presented in Tables~\ref{BF-Compare} and Tables~\ref{total-BF}.

    With $\mathcal{B}(\bar{K}^{*}(892)^{0} \rightarrow K^{-}\pi^{+})$ and $\mathcal{B}(\phi(1020) \rightarrow K^{+}K^{-})$ from PDG~\cite{PDG}, we obtain $\mathcal{B}(D_{s}^{+} \rightarrow \bar{K}^{*}(892)^{0}K^{+}) = (3.94\ \pm\ 0.12)\%$ and $\mathcal{B}(D_{s}^{+} \rightarrow \phi(1020)\pi^{+}) = (4.60\ \pm\ 0.17)\%$,
which are consistent with corresponding theory predictions~\cite{PRD93-114010}.

    \begin{table}[htbp]
        \caption{Comparisons of BFs among CLEO collaboration, Belle collaboration, BaBar collaboration and this analysis.}
        \label{BF-Compare}
        \begin{center}
            \begin{tabular}{c|c}
                \hline\hline
                $\mathcal{B}$ $(D_{s}^{+} \rightarrow K^{+}K^{-}\pi^{+})(\%)$ & Collaboration  \\
                \hline
                $5.55\pm0.14_{{\rm stat}}\pm0.13_{{\rm sys}}$    &  CLEO~\cite{CLEO-BF}        \\
                $5.06\pm0.15_{{\rm stat}}\pm0.21_{{\rm sys}}$    &  Belle~\cite{BELL-BF}       \\
                $5.78\pm0.20_{{\rm stat}}\pm0.30_{{\rm sys}}$    &  BaBar~\cite{BABAR-BF}      \\
                $5.47\pm0.08_{{\rm stat}}\pm0.13_{{\rm sys}}$    &  BESIII(this analysis)      \\
                \hline\hline
            \end{tabular}
        \end{center}
    \end{table}
    \begin{table*}[htbp]
        \caption{The BFs measured in this analysis and quoted from PDG~\cite{PDG}.}
        \renewcommand\arraystretch{1.2}
        \label{total-BF}
        \begin{center}
            \begin{tabular}{l|cc}
                \hline\hline
                \multirow{2}{*}{Process} & \multicolumn{2}{c}{BF (\%)}\\
                 &                  BESIII(this analysis)  & PDG \\
                \hline
                $D_{s}^{+} \rightarrow \bar{K}^{*}(892)^{0}K^{+}$, $\bar{K}^{*}(892)^{0} \rightarrow K^{-}\pi^{+}$              & $2.64\ \pm\ 0.06_{{\rm stat}}\ \pm\ 0.07_{{\rm sys}}$  &\hphantom{00}2.58$\ \pm \ $0.08\hphantom{00}          \\
                $D_{s}^{+} \rightarrow \phi(1020)\pi^{+}$, $\phi(1020) \rightarrow K^{+}K^{-}$                                  & $2.21\ \pm\ 0.05_{{\rm stat}}\ \pm\ 0.07_{{\rm sys}}$  &\hphantom{00}2.24$\ \pm \ $0.08\hphantom{00}        \\
                $D_{s}^{+} \rightarrow S(980)\pi^{+}$, $S(980) \rightarrow K^{+}K^{-}$                                          & $1.05\ \pm\ 0.04_{{\rm stat}}\ \pm\ 0.06_{{\rm sys}}$  &\hphantom{00}1.14$\ \pm \ $0.31\hphantom{00}       \\
                $D_{s}^{+} \rightarrow \bar{K}^{*}_{0}(1430)^{0}K^{+}$, $\bar{K}^{*}_{0}(1430)^{0} \rightarrow K^{-}\pi^{+}$    & $0.16\ \pm\ 0.03_{{\rm stat}}\ \pm\ 0.03_{{\rm sys}}$  &\hphantom{00}0.18$\ \pm \ $0.04\hphantom{00}      \\
                $D_{s}^{+} \rightarrow f_{0}(1710)\pi^{+}$ ,$f_{0}(1710) \rightarrow K^{+}K^{-}$                                & $0.10\ \pm\ 0.02_{{\rm stat}}\ \pm\ 0.03_{{\rm sys}}$  &\hphantom{00}0.07$\ \pm \ $0.03\hphantom{00}     \\
                $D_{s}^{+} \rightarrow f_{0}(1370)\pi^{+}$ ,$f_{0}(1370) \rightarrow K^{+}K^{-}$                                & $0.07\ \pm\ 0.02_{{\rm stat}}\ \pm\ 0.01_{{\rm sys}}$  &\hphantom{00}0.07$\ \pm \ $0.05\hphantom{00}    \\
                $D_{s}^{+} \rightarrow K^{+}K^{-}\pi^{+}$ total BF                                                              & $5.47\ \pm\ 0.08_{{\rm stat}}\ \pm\ 0.13_{{\rm sys}}$  &\hphantom{00}5.39$\ \pm \ $0.15\hphantom{00} \\
                \hline\hline
        \end{tabular}
    \end{center}
\end{table*}

\vspace{3.5cm}
\begin{acknowledgements}
    \label{sec:acknowledgement}
    The BESIII collaboration thanks the staff of BEPCII and the IHEP computing center for their strong support. This work is supported in part by National Key Research and Development Program of China under Contracts No. 2020YFA0406400 and No. 2020YFA0406300; National Natural Science Foundation of China (NSFC) under Contracts Nos. 11625523, 11635010, 11735014, 11822506, 11835012, 11935015, 11935016, 11935018, 11961141012; the Chinese Academy of Sciences (CAS) Large-Scale Scientific Facility Program; Joint Large-Scale Scientific Facility Funds of the NSFC and CAS under Contracts Nos. U1732263, U1832207; CAS Key Research Program of Frontier Sciences under Contracts Nos. QYZDJ-SSW-SLH003, QYZDJ-SSW-SLH040; 100 Talents Program of CAS; INPAC and Shanghai Key Laboratory for Particle Physics and Cosmology; ERC under Contract No. 758462; German Research Foundation DFG under Contracts Nos. 443159800, Collaborative Research Center CRC 1044, FOR 2359, FOR 2359, GRK 214; Istituto Nazionale di Fisica Nucleare, Italy; Ministry of Development of Turkey under Contract No. DPT2006K-120470; National Science and Technology fund; Olle Engkvist Foundation under Contract No. 200-0605; STFC (United Kingdom); The Knut and Alice Wallenberg Foundation (Sweden) under Contract No. 2016.0157; The Royal Society, UK under Contracts Nos. DH140054, DH160214; The Swedish Research Council; U. S. Department of Energy under Contracts Nos. DE-FG02-05ER41374, DE-SC-0012069
\end{acknowledgements}



\end{document}